\documentclass[11pt,a4paper]{article}

\usepackage[top=2cm, bottom=2cm, left=2cm, right=2cm]{geometry}
\usepackage[nodisplayskipstretch]{setspace}
\setdisplayskipstretch{0.3}

\usepackage[subtle]{savetrees}

\usepackage[compact]{titlesec}

\usepackage{caption}
\usepackage{natbib}
\setcitestyle{authoryear, aysep={}, yysep={;}}

\usepackage[utf8]{inputenc}
\usepackage{amsmath}
\usepackage{amsthm}
\theoremstyle{plain}
\usepackage{amsfonts}
\usepackage{mathtools}
\usepackage{dsfont}
\usepackage{graphicx}
\usepackage{nicefrac}
\usepackage{bm}
\usepackage{pbox}
\usepackage{makecell}
\usepackage{physics}
\usepackage[dvipsnames]{xcolor}

\usepackage[shortlabels]{enumitem}
\setitemize{noitemsep,topsep=10pt,parsep=0pt,partopsep=0pt}
\setenumerate{noitemsep,topsep=10pt,parsep=0pt,partopsep=0pt}

\usepackage{placeins}
\usepackage{authblk}
\usepackage{array}
\usepackage{booktabs}
\usepackage{ragged2e}

\usepackage[hidelinks]{hyperref}

\graphicspath{ {./paper_figures/} }

\newtheorem{proposition}{Proposition}

\newcommand{\bmat}[1]{
  \renewcommand*{\arraystretch}{0.8}
  \begin{bmatrix}#1\end{bmatrix}%
}

\title{Harmonized Estimation of Subgroup-Specific Treatment Effects in Randomized Trials: The Use of External Control Data}

\author[1,3,*]{Daniel Schwartz}
\author[1]{Riddhiman Saha}
\author[2]{Steffen Ventz}
\author[1,3]{Lorenzo Trippa}
\affil[1]{Department of Biostatistics, Harvard T.H. Chan School of Public Health, United States}
\affil[2]{Division of Biostatistics, School of Public Health, University of Minnesota, United States}
\affil[3]{Department of Data Science, Dana-Farber Cancer Institute, United States}
\affil[*]{Corresponding author: daniels@ds.dfci.harvard.edu}

\date{January 29, 2025}

\begin{document}

\maketitle

\begin{abstract}
    Subgroup analyses of randomized controlled trials (RCTs) constitute an important component of the drug development process in precision medicine. In particular, subgroup analyses of early-stage trials often influence the design and eligibility criteria of subsequent confirmatory trials and ultimately influence which subpopulations will receive the treatment after regulatory approval. However, subgroup analyses are often complicated by small sample sizes, which leads to substantial uncertainty about subgroup-specific treatment effects. We explore the use of external control (EC) data to augment RCT subgroup analyses. We define and discuss {\it harmonized estimators } of subpopulation-specific treatment effects that leverage EC data. Our approach can be used to modify any subgroup-specific treatment effect estimates that are obtained by combining RCT and EC data, such as linear regression. We alter these subgroup-specific estimates to make them coherent with a robust estimate of the average effect in the randomized population based only on RCT data. The weighted average of the resulting subgroup-specific harmonized estimates matches the RCT-only estimate of the overall effect in the randomized population. We discuss the proposed harmonized estimators through analytic results and simulations, and investigate standard performance metrics. The method is illustrated with a case study in oncology.
    
    \end{abstract}


\section{Introduction}

Evaluating the efficacy of experimental therapies in subpopulations defined by biomarkers that modulate the therapeutic mechanisms has become a major goal of precision medicine. However, the sample size of randomized controlled trials (RCTs) is typically chosen with the goal of estimating the therapy's average effect in the  randomized population. Consequently, RCT sample sizes are often insufficient to accurately evaluate treatment effects in the subpopulations of interest. Yet subgroup analyses are often  reported despite limited sample sizes. In early-phase trials these analyses are important for major decisions, such as determining the design and eligibility criteria for subsequent confirmatory trials \citep{sun_credibility_2012, pletcher_challenges_2017}. In late phase trials, if the therapy receives regulatory approval, subgroup analyses may influence how clinicians choose individualized treatment plans for their patients \citep{dijkman_how_2009}. Unfortunately, limited resources, time, and patient accrual rates often make it infeasible to conduct large trials that can accurately estimate treatment effects within each subpopulation \citep{sully_reinvestigation_2013}.

To address this limitation, we investigate potential improvements of  RCT subgroup analyses by including external control (EC) data from patients that were treated with the RCT's control therapy.  We consider  EC data from previous trials or electronic health records. At completion of the RCT we augment the subgroup analyses with patient-level external control data, including relevant pre-treatment clinical and demographic variables, and  outcomes. Here, we introduce a method  that balances the potential efficiency gains from the integration of external data and the risk of bias from unmeasured confounding or other distortion mechanisms.


Our work is motivated by two oncology trials in brain and breast cancer that differ substantially in terms of sample sizes and prevalence of the subgroups of interest. In both cases, subgroup analyses using only internal RCT data result in high uncertainty on the subgroup-specific treatment effects. This level of uncertainty affects several decisions, including the choice of designs and eligibility criteria (i.e., selected subgroups) of future confirmatory trials. 

\begin{enumerate}[(a)]
\item \textit{Temozolomide for glioblastoma.} A large RCT in glioblastoma (GBM), a brain cancer with poor prognosis,  showed that \textit{temozolomide} used in combination with radiotherapy (TMZ+RT) is an effective first line treatment compared to radiotherapy alone \citep{stupp_radiotherapy_2005}. As of today TMZ+RT remains the standard of care for GBM. After the RCT results were reported, the study investigators collected MGMT methylation data, an important biomarker in neuro-oncology.
 Subgroup analyses suggested that the treatment effect of TMZ+RT compared to RT alone may vary by methylation status, with methylation-positive patients benefiting from TMZ+RT and high uncertainty about whether or not methylation-negative patients  benefit from the combination therapy \citep{stupp_effects_2009}. The sample size of the biomarker negative subgroup was not sufficient to provide definitive evidence about treatment efficacy in this subpopulation. 
     
\item \textit{Olapirib for metastatic breast cancer.} Olapirib is a promising drug approved to treat metastatic breast cancer. 
A recent RCT tested its efficacy in two populations defined by  sets  of somatic (population 1) and germline  (population 2) mutations. We refer to \cite{tung_tbcrc_2020} for details. The trial was powered to test efficacy in these two populations. The analyses indicated that the benefit of olapirib might be limited to some of the patients in populations 1 and 2 with 
specific mutations.
 However, the study had limited power to test efficacy in narrow subgroups defined by  the individual mutations without grouping them into the broad categories of somatic and germline alterations.
\end{enumerate}
Our analyses (Section \ref{section:GBM}) are based on a recent data collection in glioblastoma \citep{rahman_accessible_2023}. The aim is to illustrate the operating characteristics of subgroup-specific treatment effect estimators with comparisons tailored to a specific clinical setting.

Most methods for subgroup analyses use only internal data from the RCT. The simplest and perhaps most common approach in clinical research is to repeat the primary analysis, for instance a  two-sample Z-test, in each of the subgroups of interest \citep{wang_statistics_2007}. 
This strategy is often underpowered for testing whether or not the experimental treatment is effective in the subgroups of interest \citep{pletcher_challenges_2017}. Regression methods that model the outcome conditional on biomarkers and other covariates have been proposed to reduce the uncertainty of subgroup-specific results \citep{morita_bayesian_2017, sivaganesan_subgroup_2017}. 

The use of EC data for inference at completion of RCTs has been previously discussed \citep{corrigan-curay_real-world_2018, di_maio_real-world_2020}. However, these approaches leverage RCT and EC data primarily to estimate average treatment effects in the \textit{randomized} population, not in subpopulations. Bayesian methods in this area include meta-analytic models \citep{schmidli_robust_2014}, power priors \citep{ibrahim_power_2015}, and commensurate priors \citep{hobbs_commensurate_2012}. Other approaches include the use of propensity scores \citep{lim_minimizing_2018, li_generalizing_2022} and adaptive procedures that compare the RCT control data with the EC data to decide whether to pool them in the analyses or use only the RCT data for inference \citep{viele_use_2014}. EC data have also been used for sample size re-estimation \citep{hobbs_adaptive_2013, schmidli_robust_2014} and early stopping for futility \citep{ventz_use_2022}. 

We focus on patient subgroups and potential treatment effects' variations across subgroups. We investigate if incorporating EC data can produce a substantial reduction of the uncertainty on the subgroup-specific treatment effects. 
At the same time, we account for the fact that most RCTs are required by their protocol to report an estimate $\hat{\theta}^{(r)}$ of the average treatment effect that is based only on trial data and \textit{does not} incorporate EC data. This estimate and any subgroup-specific estimate $\hat{\theta}^{(r+e)}_{1:K}$ based on  both RCT and EC data could be discordant, for example  $\hat{\theta}^{(r)}$ might be negative and the subgroup-specific estimates could be all positive. This discordance can arise because $\hat{\theta}^{(r)}$ and $\hat{\theta}^{(r+e)}_{1:K}$ are based on different sets of data. We introduce harmonization to reduce or remove this misalignment.

Our method requires two inputs: (i) an  estimate of the average treatment effect in the randomized population $\hat{\theta}^{(r)}$, which is based only on the RCT data, and (ii) an estimate of subgroup-specific effects $\hat{\theta}^{(r+e)}_{1:K}$ in $K$ non-overlapping subgroups, which is based on both RCT and EC data, with some risk of bias. We use the superscript $(r+e)$ for estimators that use both RCT and EC data,  and $(r)$ for estimators that use only RCT data. 
The method is based on a simple linear transformation (equation \eqref{harmonized_simple}) to modify the subgroup-specific estimates $\hat{\theta}^{(r+e)}_{1:K}$ and harmonize them (i.e., make them coherent) with respect to the average treatment effect estimate $\hat{\theta}^{(r)}$.
The term \textit{harmonization} refers to an intuitive relation between the  average treatment effect $E \left( Y | T = 1 \right) - E \left( Y | T = 0 \right)$ in the trial population, and the  treatment effects $E \left( Y | T = 1, W = k \right) - E \left( Y | T = 0, W = k \right)$ in the  subgroups  $k=1,\ldots,K$ that partition the randomized population:
\begin{equation*}
    E \left( Y | T = 1 \right) - E \left( Y | T = 0 \right) = \sum_{k=1}^K \pi_k \Big[ E \left( Y | T = 1, W = k \right) - E \left( Y | T = 0, W = k \right) \Big].
\end{equation*}
Here $W$ indicates the subpopulation and $\pi_1, \dots, \pi_K$ are the subpopulation prevalences in the trial population. In the RCT patients are randomized and the variables $T$ and $W$ are independent.
We say that a vector of subgroup-specific  estimates $\hat{\theta}_{1:K}$ is fully harmonized to the primary analysis if $\sum_{k=1}^K \pi_k \hat{\theta}_k = \hat{\theta}^{(r)}$. When a subgroup analysis that leverages ECs is not harmonized and there is a marked discrepancy between $\sum_{k=1}^K \pi_k \hat{\theta}_k^{(r+e)} $ and $  \hat{\theta}^{(r)}$ it can be challenging to interpret the  trial results.

Our method  modifies initial subgroup-specific estimates $\hat{\theta}^{(r+e)}_{1:K}$ 
and harmonizes them to $\hat{\theta}^{(r)}$; therefore the subgroup analyses become coherent with a primary analysis that reports $\hat{\theta}^{(r)}$ for the randomized population without using EC data. 
We illustrate that the  subgroup-specific harmonized estimates present an attractive balance between efficiency and bias in settings where the investigator expects  identical or similar differences $\gamma_k \approx \gamma$  between the  outcomes in the EC population  and the  outcomes in the RCT control arm across all subgroups $k=1,\ldots, K$. Formal definitions of these discrepancies    $\gamma_k$ are provided in the next sections.

\section{Harmonized estimates of subgroup-specific treatment effects}

We consider a two-arm RCT. The analysis of the trial can leverage EC data of patients receiving the same therapy as in the RCT control group. Before treatment patients are categorized into non-overlapping  biomarker subgroups $k = 1, \dots, K$. 
When  necessary we use the superscript $(r)$ to indicate  the RCT samples, $(e)$ to denote the ECs, and $(r+e)$ for their union. For example the  sample sizes are $n^{(r)}, n^{(e)}$ and $n^{(r+e)}$. Also, the subgroup- and treatment-specific sample sizes are $n_{k,t}^{(s)}$, where $t = 0$ for the control, $t = 1$ for the experimental treatment, and $s \in \{r, e, r+e\}$. Similarly, $n_{k,\cdot}^{(r)}$
 is the total number of RCT patients in subgroup $k$ and $n_{\cdot,0}^{(r+e)}$ is the  number of control patients in the RCT and EC datasets. The notation is summarized in Table \ref{notation_table}. 

For the $i$-th patient in the RCT we have the vector $\left( T_i^{(r)}, W_i^{(r)}, Y_i^{(r)}, X_i^{(r)} \right)$, where $T_i^{(r)} \in \{0,1\}$ is the treatment indicator (0 for control, 1 for treatment), $W_i^{(r)} \in \{1, \dots, K\}$ is the biomarker subgroup, $Y_i^{(r)}$ is the outcome, and $X_i^{(r)}$ is a vector of pre-treatment covariates. Similarly, for the $i$-th patient in the EC data we have $\left( T_i^{(e)}, W_i^{(e)}, Y_i^{(e)}, X_i^{(e)} \right)$, with $T_i^{(e)} = 0$ because all patients in the external dataset received the control therapy. The RCT  and  EC datasets are indicated as $D^{(r)}$ and $D^{(e)}$ respectively. 

In the RCT population, within subgroup $k$ the expected outcome under the control therapy is $\mu_k = E \left( Y_i^{(r)} | W_i^{(r)} = k, T_i^{(r)} = 0 \right)$, and the treatment effect is $$\theta_k = E \left( Y_i^{(r)} | W_i^{(r)} = k, T_i^{(r)} = 1 \right) - E \left( Y_i^{(r)} | W_i^{(r)} = k, T_i^{(r)} = 0 \right).$$ The expected outcome and treatment effect in the  RCT population are $\mu = E \left( Y_i^{(r)} | T_i^{(r)} = 0 \right) $ and $$\theta = E \left( Y_i^{(r)} | T_i^{(r)} = 1 \right) - E \left( Y_i^{(r)} | T_i^{(r)} = 0 \right) = \sum_{k=1}^K \pi_k \theta_k,$$ where $\pi_k = p \left( W_i^{(r)} = k \right)$ is the prevalence of the subgroup $k$ in the trial population.

\begin{table}[htbp!]

\centering

\renewcommand{\arraystretch}{1.4}

\begin{tabular}{>{\raggedright\hangindent=1em}m{0.3\linewidth} m{0.22\linewidth} m{0.40\linewidth}}

\toprule
\multicolumn{1}{l}{Variable} & \multicolumn{1}{c}{Notation} & \multicolumn{1}{c}{Definition} \\ 
\hline
Outcome & $Y_i$ &  \\
Treatment & $T_i$ & 0 for control arm, 1 for treatment arm \\
Study &  & $(r)$ for RCT, $(e)$ for EC \\
Subgroup & $W_i$ & $W_i = k$ if patient $i$ is in subgroup $k$ \\
Covariates & $X_{i}$ & Vector of pre-treatment covariates   \\
Datasets & $D^{(r)}, D^{(e)}$ & $(T_i, W_i, Y_i, X_i)$ for all RCT and EC patients respectively \\
Group-specific sample sizes & $n_{k,t}^{(s)}$ & In subgroup $k$, arm $t$, and study $s$ \\
Subgroup prevalences in the RCT population & $\pi = (\pi_1, \dots, \pi_K)$ & $\pi_k = p \left( W_i^{(r)} = k \right)$ \\
Expected control outcomes  in the  RCT subgroups & $\mu_{1:K}\! = \!(\mu_1, \dots, \mu_K)$ & $ \mu_k = E \left( Y_i^{(r)} | W_i^{(r)} = k, T_i^{(r)} = 0 \right)$ \\
Expected control outcomes in the RCT population & $\mu$ & $ E \left( Y_i^{(r)} | T_i^{(r)} = 0 \right) = \sum_{k = 1}^K \pi_k \mu_k$ \\
Treatment effects in the RCT subpopulations & $\theta_{1:K} = (\theta_1, \dots, \theta_K)$ & {$\begin{aligned}
    \theta_k & = E \left( Y_i^{(r)} | W_i^{(r)} = k, T_i^{(r)} = 1 \right) - \mu_k
\end{aligned}$} \\
Treatment effect in the RCT population & $\theta$ & $E \left( Y_i^{(r)} | T_i^{(r)} = 1 \right) - \mu = \sum_{k=1}^K \pi_k \theta_k$ \\
Initial estimator of $\theta_{1:K}$ & $\hat{\theta}^{(r+e)}_{1:K}$ & A function of  the RCT data $D^{(r)}$ and the EC data $D^{(e)}$ \\
Estimator of $\theta$ & $\hat{\theta}^{(r)}$ & A function of  the RCT    data $D^{(r)}$ \\
Harmonized estimator of $\theta_{1:K}$ & $\hat{\theta}_{1:K}^h$ & Defined by equation \eqref{harmonized_def} \\

\bottomrule
\end{tabular}
\textbf{}
\caption{Notation.}

\label{notation_table}

\end{table}

\subsection{Harmonized estimates}

The subgroup-specific treatment effect estimates that we propose are harmonized to the  primary analysis, which utilizes only   RCT data. The definition of the harmonized estimator proceeds in two steps. 
In the first step we compute an initial estimate of the subgroup-specific treatment effects $\hat{\theta}_{1:K}^{(r+e)}$, using both the RCT data $D^{(r)}$ and the EC data $D^{(e)}$. 
The analyst  chooses the estimator $\hat{\theta}^{(r+e)}_{1:K}$.
 For example, $\hat{\theta}_{1:K}^{(r+e)}$ can be the result of a regression analysis    
   that   merges $D^{(r)}$ and $D^{(e)}$ into a single dataset    
   under the  assumption  that the RCT and EC populations are identical, or the primary findings of a Bayesian analysis that  models the RCT and EC populations. 
 Throughout the paper we consider several definitions of $\hat{\theta}_{1:K}^{(r+e)}$. 
 In the second step we modify $\hat{\theta}_{1:K}^{(r+e)}$ and harmonize it with respect to an overall treatment effect estimate  $\hat{\theta}^{(r)}$ based only on $D^{(r)}$. 
 Here $\hat{\theta}^{(r)}$ is an estimate associated with low risks of bias, confounding, and various distortion mechanisms, and is based only on RCT data. We say that a vector of estimates $\hat{\theta}_{1:K}$ is fully harmonized if $\sum_{k=1}^K \pi_k \hat{\theta}_k = \hat{\theta}^{(r)}$, i.e. if its weighted average  equals the  estimate of the average treatment effect. Recall that $\pi = (\pi_1, \dots, \pi_K)$ are the subgroup prevalences in the RCT population. The harmonized estimator is defined as
  \begin{equation} \label{harmonized_def}
    \hat{\theta}_{1:K}^{h} = \underset{v \in \mathbb{R}^K}{
    \text{arg min}} \left[ \left( v - \hat{\theta}_{1:K}^{(r+e)} \right)^\top \Sigma^{{-1}} \left( v - \hat{\theta}_{1:K}^{(r+e)} \right) + \lambda \left( \pi^\top v - \hat{\theta}^{(r)} \right)^2 \right],
\end{equation}
where $\Sigma$ is a positive definite $K \times K$ matrix and $\lambda$ is a non-negative parameter that controls how closely the weighted average $\sum_{k=1}^K \pi_k \hat{\theta}_k^{h}$ matches the RCT estimate $\hat{\theta}^{(r)}$. When $\lambda = 0$ the harmonized estimator is equal to $\hat{\theta}^{(r+e)}_{1:K}$, while when $\lambda = \infty$ it becomes fully harmonized and $\sum_{k=1}^K \pi_k \hat{\theta}_k^{h} = \hat{\theta}^{(r)}$. Our paper examines the  class of harmonized estimators defined by equation  (\ref{harmonized_def}), with particular attention to harmonized estimators with $\lambda = \infty$.  

As a simple example of  harmonized estimator, consider the following:
\begin{enumerate}[wide = 0.5\parindent, leftmargin = 0.5\parindent]
    \item[\textit{(Step 1)}]  Use the pooled estimates
    \begin{equation} \label{eq:basic_subgroup}
        \hat{\theta}^{(r+e)}_k  = \bar{Y}_{k, 1}^{(r)} - \bar{Y}_{k, 0}^{(r+e)}, 
    \end{equation}
    for each subgroup $k$. Here $\bar{Y}_{k, 1}^{(r)}$ is the mean outcome in subgroup $k$ under the experimental treatment, and $\bar{Y}_{k, 0}^{(r+e)}$ is the mean outcome under the control therapy, including both the EC data $D^{(r)}$ and the external data $D^{(e)}$.

    \item[\textit{(Step 2)}] The estimate of the treatment effect in the RCT population is
    \begin{equation} \label{eq:basic_overall}
        \hat{\theta}^{(r)} = \bar{Y}_{\cdot,1}^{(r)} - \bar{Y}_{\cdot,0}^{(r)},
    \end{equation}
    the difference between the mean outcomes in the experimental and control RCT arms, excluding the EC data. In this  example if $\Sigma$ is the identity matrix then $\left( v - \hat{\theta}_{1:K}^{(r+e)} \right)^\top \Sigma^{{-1}} \left( v - \hat{\theta}_{1:K}^{(r+e)} \right)$ in \eqref{harmonized_def} is the squared Euclidean distance between $\hat{\theta}^{(r+e)}_{1:K}$ and $v$. Similarly, if $\Sigma$ is  an estimate of the covariance matrix of $\hat{\theta}^{(r+e)}_{1:K}$, then the first term in the right part of  equation \eqref{harmonized_def} becomes the squared Mahalanobis distance.
\end{enumerate}

We mention two practical aspects of the harmonized estimator. First, it ensures that the subgroup and primary analyses provide coherent inference about treatment effects. Indeed,  we infer subgroup-specific treatment effects with a weighted average close to the point estimate $\hat{\theta}^{(r)}$ in the primary analysis. Other methods that leverage both $D^{(r)}$ and $D^{(e)}$  (e.g.,  expression \ref{eq:basic_subgroup}) may produce subgroup-specific results that are clearly incongruous  with $\hat{\theta}^{(r)}$. Second, a broad range of  estimators $\hat{\theta}^{(r+e)}_{1:K}$ can be used, for example: 
\begin{enumerate}[(a), wide=0.5\parindent, leftmargin = 0.5\parindent]
    \item the simple difference between $\bar{Y}_{k,1}^{(r+e)}$ and $\bar{Y}_{k,0}^{(r+e)}$ as in equation \eqref{eq:basic_subgroup},
    \item the results of a Bayesian analysis based on a flexible regression model where the outcome distribution varies across  biomarker subgroups, treatments, populations (RCT or EC), and patient-level pre-treatment covariates \citep[e.g.,][]{sivaganesan_subgroup_2017},
    \item or estimates based on a matching  procedure \citep[e.g.,][]{lin_propensityscorebased_2019} applied within each subgroup, to account for different distributions of potential confounders in the RCT and EC populations.
\end{enumerate}

The solution of equation \eqref{harmonized_def} is
\begin{equation} \label{harmonized_shrinkage}
    \hat{\theta}_{1:K}^{h} = \left( \Sigma^{-1} + \lambda \pi \pi^\top \right)^{-1} \left[ \Sigma^{-1} \hat{\theta}^{(r+e)}_{1:K} + \lambda \pi \pi^\top \left( \hat{\theta}^{(r)} 1_{K \times 1} \right) \right],
\end{equation}
where the column vector $1_{K \times 1} = (1,  \dots, 1)$ has length $K$. 
This equation shows that the harmonized estimator is a {\it matrix weighted average } (see \cite{chamberlain_matrix_1976}) of $\hat{\theta}^{(r+e)}_{1:K}$ and $\hat{\theta}^{(r)} 1_{K \times 1}$. The solution can be derived based on the derivatives of the objective function in \eqref{harmonized_def}
\begin{align*}
    \Sigma^{-1} v - \Sigma^{-1} \hat{\theta}^{(r+e)}_{1:K} + \lambda \pi^\top v \pi - \lambda \hat{\theta}^{(r)} \pi,   \;\; 
\end{align*} which are equal to zero when $$
    \left( \Sigma^{-1} + \lambda \pi \pi^\top \right) v = \Sigma^{-1} \hat{\theta}^{(r+e)}_{1:K} + \lambda \hat{\theta}^{(r)} \pi . $$ 
Moreover, the harmonized estimator can be rewritten as
\begin{equation} \label{harmonized_simple}
        \hat{\theta}_{1:K}^{h} = \hat{\theta}_{1:K}^{(r+e)} + c \frac{\lambda}{\lambda + c} \left( \hat{\theta}^{(r)} - 
        %
        \sum_{k=1}^K \pi_k \hat{\theta}_k^{(r+e)}
         \right) \Sigma \pi,
\end{equation}
where $c = \left( \pi^\top \Sigma \pi \right)^{-1}$.  
This representation of $\hat{\theta}^h_{1:K}$ shows that the harmonized estimator is obtained by shifting  $\hat{\theta}^{(r+e)}_{1:K}$ in a fixed direction, $\Sigma \pi$. The distance  
between  $ \hat{\theta}^{h}_{1:K} $ and $\hat{\theta}^{(r+e)}_{1:K}$
 is proportional to the discrepancy between $\hat{\theta}^{(r)}$ and $\pi^\top \hat{\theta}_{1:K}^{(r+e)}$. For example, if $\lambda = \infty$ and $\Sigma \pi \propto 1_{K \times 1}$ then in each subgroup $\hat{\theta}^{h}_{k} - \hat{\theta}^{(r+e)}_{k} = \hat{\theta}^{(r)} - \pi^\top \hat{\theta}_{1:K}^{(r+e)}$. 
We derived  equation \eqref{harmonized_simple} using the inversion formula $(\Sigma^{-1} + \lambda \pi \pi^\top)^{-1} = \Sigma - 
\lambda
(1 + \lambda \pi^\top \Sigma \pi)^{-1}
 \Sigma \pi \pi^\top \Sigma$.

{\it The choice of $\Sigma$.}
In this paper we discuss  two  approaches for  selecting  $\Sigma$, which lead to what we call bias-directed (BD) harmonization and variance-directed (VD) harmonization. 

{\it BD harmonization:} In several cases $\Sigma$ can be chosen by leveraging an estimate of the bias of $\hat{\theta}_{1:K}^{(r+e)}$, which we denote as $\psi_{1:K} = E \left( \hat{\theta}^{(r+e)}_{1:K} \right) - \theta_{1:K}$. See for example expression \eqref{relaxed_bias_variance} in the next subsection. Harmonization with $\lambda = \infty$ and a $\Sigma$ matrix chosen to satisfy $\Sigma \pi \propto \psi_{1:K}$ provides unbiased estimates of $\theta_{1:K}$. We refer to  $\hat{\theta}_{1:K}^h$  with this choice of $\Sigma $ and $\lambda$ as the \textit{BD harmonized estimator}. Indeed, when $\lambda = \infty$ and $\Sigma \pi \propto \psi_{1:K}$, equation \eqref{harmonized_simple} can be rewritten as
\begin{equation*}
    \hat{\theta}^h_{1:K} = \hat{\theta}^{(r+e)}_{1:K} - \left( \pi^\top \hat{\theta}^{(r+e)}_{1:K} - \hat{\theta}^{(r)} \right) \left( \pi^\top \Sigma \pi \right)^{-1} \Sigma \pi,
\end{equation*}
and the second term on the right hand side is an  unbiased estimate of  $\psi_{1:K}$. We note that, if $E \left( \hat{\theta}^{(r)} \right) = \theta = \pi^\top \theta_{1:K}$ holds, then $E \left( \pi^\top \hat{\theta}^{(r+e)}_{1:K} - \hat{\theta}^{(r)} \right) = \pi^\top \psi_{1:K}$ and 
\begin{equation*}
    E \left[ \left( \pi^\top \hat{\theta}^{(r+e)}_{1:K} - \hat{\theta}^{(r)} \right) \left( \pi^\top \Sigma \pi \right)^{-1} \Sigma \pi \right] = \pi^\top \psi_{1:K} \left( \pi^\top \psi_{1:K} \right)^{-1} \psi_{1:K} = \psi_{1:K}.
\end{equation*}
We will use similar arguments in later sections to derive BD harmonized estimators when the input $\hat\theta^{(r+e)}_{1:K}$ is based on regression models. For example, in Section \ref{subsection:logistic_model}, we consider logistic regression.

{\it VD harmonization:} Another interpretable choice  is  $\Sigma$ proportional to an estimate of the  covariance matrix of $\hat{\theta}^{(r+e)}_{1:K}$. When  $\lambda = \infty$, we refer to $\hat{\theta}_{1:K}^h$  as the \textit{VD harmonized estimator}.  A Bayesian interpretation is discussed in Section \ref{subsection:cut}.

\subsection{Bias and variance of the harmonized estimator} \label{subsection:simple_bv}

When the joint distribution of $\hat{\theta}_{1:K}^{(r+e)}$ and $\hat{\theta}^{(r)}$ is known we can derive the bias and variance of $\hat{\theta}^h_{1:K}$, for any choice of $\lambda$ and $\Sigma$, because the harmonized estimator is a linear function of $\hat{\theta}_{1:K}^{(r+e)}$ and $\hat{\theta}^{(r)}$ (equation \eqref{harmonized_simple}). In this subsection we specify scenarios (i.e.,  data-generating models), consider simple estimators  $\hat{\theta}_{1:K}^{(r+e)}$ and $\hat{\theta}^{(r)}$, and  illustrate the bias, variance and MSE of $\hat{\theta}^h_{1:K}$ based on  analytic expressions. In Sections 3 and 4 we will discuss harmonization in other settings, with different scenarios and definitions of  $\hat{\theta}_{1:K}^{(r+e)}$ and $\hat{\theta}^{(r)}$.

We consider data generated from the following  model:
\begin{equation} \label{simple_model}
    \begin{split}
        Y_i^{(r)} | W_i^{(r)}, T_i^{(r)} &\overset{iid}{\sim}  N \left( \mu_{W_i^{(r)}} + \theta_{W_i^{(r)}} T_i^{(r)}, \phi^2 \right), \text{ and} \\
       Y_i^{(e)} | W_i^{(e)} &\overset{iid}{\sim} N \left(  \mu_{W_i^{(e)}} + \gamma_{W_i^{(e)}}, \phi^2 \right).
    \end{split}
\end{equation}
Here $\mu_k$ is the expected outcome under the control therapy for subgroup $k$ in the RCT population, $\theta_k$ is the subgroup-specific treatment effect, and $\gamma_k$ is a bias term, the difference between the expected outcomes of the external and RCT controls in subgroup $k$. All patients in the external data receive the control therapy. 
Model \eqref{simple_model} does not include covariates $X$.

As inputs for $\hat{\theta}^h_{1:K}$, we consider the   estimators $\hat{\theta}_{k}^{(r+e)} = \bar{Y}_{k, 1}^{(r)} - \bar{Y}_{k, 0}^{(r+e)}$ and $ \hat{\theta}^{(r)} = \bar{Y}_{\cdot,1}^{(r)} - \bar{Y}_{\cdot,0}^{(r)}$ as in equations \eqref{eq:basic_subgroup} and \eqref{eq:basic_overall}. For the study design, we assume that the vector $\pi$ is known and that for each subgroup $k$
$$  \frac{n_{k,0}^{(r)}}{n_{\cdot,0}^{(r)}} = \frac{n_{k,1}^{(r)}}{n_{\cdot,1}^{(r)}} = \pi_k. $$
This condition is  satisfied if recruitment is stratified by subgroup and the trial adopts block randomization. For simplicity, we also assume the same proportions in the EC dataset, i.e. for each subgroup $k$
$$ \frac{n_{k,\cdot}^{(e)}}{n^{(e)}} = \pi_k.$$
After Proposition \ref{prop:simple}, all other theoretical results in the paper will not assume equal  subgroup proportions in the RCT  and external controls, see for example the last paragraphs of this subsection where we consider settings with $\frac{n_{k,0}^{(e)}}{n^{(e)}} \neq \pi_k$.

We derive the sampling distribution of the harmonized estimator in this setting. We first restrict attention to the case where randomization 
is  balanced, i.e. $n_{\cdot,1}^{(r)} = n_{\cdot,0}^{(r)}$, and $\pi = (1/K, \dots, 1/K)$. We will later relax the assumptions of this stylized example. Let $q = \frac{n^{(e)}_{\cdot, 0}}{n^{(r)}_{\cdot, 0} + n^{(e)}_{\cdot, 0}}$ be the proportion of controls who are in the external data. 
\begin{proposition} \label{prop:simple}
    In the outlined setting, if $\Sigma^{-1} = I_K$, then 
    $\hat{\theta}^{h}_{1:K} \sim N \left( \theta_{1:K} + B^h, V^h \right)$, 
    where the bias vector $B^h$ is  
    \begin{equation} \label{simple_bias}
        B^h = - q \left( I_K - \frac{1}{K} \frac{\lambda}{\lambda + K} 1_{K \times K} \right) \gamma_{1:K},
    \end{equation}
    and the variance-covariance matrix is 
    \begin{equation} \label{simple_variance}
         V^h = K \frac{1}{n_{\cdot,0}^{(r)}} (2 - q) \phi^2 I_K + \left( \frac{\lambda}{\lambda + K} \right)^2 q \frac{1}{n_{\cdot, 0}^{(r)}} \phi^2 1_{K \times K}.
    \end{equation}
\end{proposition}
Here $1_{K \times K}$ is a $K \times K$ matrix with all elements equal to 1. As expected the harmonized estimator is normally distributed; indeed it is a linear transformation of the outcome data (see equation \ref{harmonized_shrinkage}). The derivations of this proposition and  other  results  are provided in the Supplementary Material. If the outcome data in $D^{(r)}$ and $D^{(e)}$ are not normal but have means and variances identical to those in \eqref{simple_model}, then the bias and variance expressions in Proposition \ref{prop:simple} still hold.

{\it Systematic distortion mechanisms.} 
In several contexts, it is implausible that the distortion mechanisms (e.g., measurement errors) cause differences between the conditional outcome distributions in the RCT control arm and the EC arm that vary across subgroups. Throughout this paper, we assume that if there is a discrepancy between the conditional outcome distributions in the control arm of the trial and the external control group, then the distortion mechanism is systematic across subgroups. This systematic distortion mechanism (SDM) assumption is formalized by the equation
\begin{equation} \label{systematic_distortion_sec_2}
    \gamma_k = \gamma, 
\end{equation}
for all subgroups $k = 1, \dots, K$ and some $\gamma \in \mathbb{R}$. In the remainder of the paper the definition of the SDM assumption will be slightly generalized { (see  \eqref{systematic_distortion_sec_3_1} and \eqref{systematic_distortion_sec_3_2})}. The SDM assumption will be relevant to interpret  the analyses  in this paper, and it holds when the difference between the distributions of the outcomes in the EC group and the RCT control arm, conditional on pre-treatment covariates, is identical across subgroups. These differences  can be summarized by comparing conditional means (here and Section 3.1), odds ratios (Section 3.2), or other  metrics. In previous work, we identified systematic distortion mechanisms in collections of datasets in neuro-oncology \citep{rahman_accessible_2023, ventz_use_2022}. 

There are several potential causes of systematic distortions. For example, the measurement of the outcomes in the external group might include  a systematic shift, in all $K$ subgroups,  compared to the measurements in the RCT. In particular, when we consider oncology RCTs, different procedures may be used to measure tumor sizes in the trial patients and the external controls. Similarly, the definition of the first day of follow up, or the schedule of radiographic measurements can vary across studies and cause systematic discrepancies in the  assessment of response to therapy and tumor progression \citep{dabush_impact_2021}. We note that in these settings the experimental therapy might produce positive effects in one or a few of the $K$ subpopulations. Systematic distortions might also arise with subtle differences between the administration of the control therapy in the RCT and the external  group, such as different levels of adherence to the schedule of the chemoradiation plan.

When the SDM assumption is satisfied and the assumptions  of Proposition \ref{prop:simple} hold, if we set $\lambda = \infty$, then the harmonized estimator is \textit{unbiased}. Indeed $B^h$ becomes $- q \left( I_K - \frac{1}{K} 1_{K \times K} \right) \gamma 1_{K \times 1} = 0_{K \times 1}$. If the subgroups are unbalanced (i.e., $\pi_k \neq \frac{1}{K}$ for some $k$), then we can choose $\Sigma$ to produce an unbiased harmonized estimator. We provide details at the end of this subsection. 

\begin{figure}[htbp!]
    \centering
    \includegraphics[width = 0.95\textwidth]{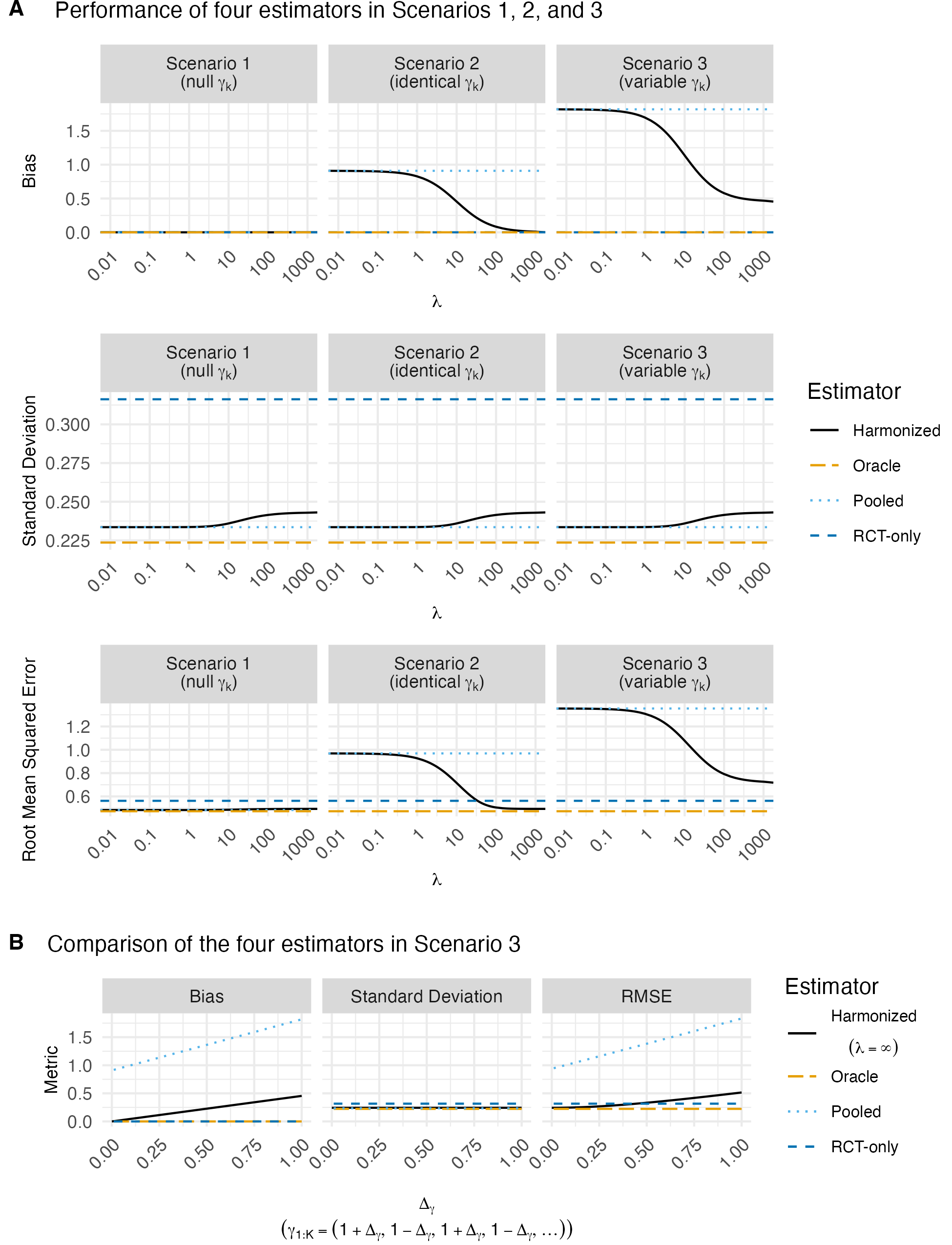}
    \caption{(A) Performance summaries of the harmonized estimator (see Proposition \ref{prop:simple}) and other estimators (analogous analytic results) for subgroup $k=1$. The $\lambda$ value  used to define the harmonized estimator varies on the x-axis. We illustrate the bias, standard deviation, and the root mean squared error (RMSE) of the estimator as a function of $\lambda$. The scenarios have different $\gamma_{1:K}$ values: $0_{K \times 1}$ (Scenario 1), constant across subgroups (Scenario 2), or  varying across subgroups, with $\gamma_{1:K} = (1 + \Delta_\gamma, 1 - \Delta_\gamma, 1 + \Delta_\gamma, 1 - \Delta_\gamma, \dots)$ and $\Delta_\gamma = 1$ (Scenario 3). (B) The performance of the harmonized estimator (with $\lambda = \infty$) for subgroup $k=1$ degrades when the subgroup-specific differences $\gamma_{1:K}$ vary substantially across subgroups, i.e. as $\Delta_\gamma$ grows.}
    \label{fig:fig_1_new}
\end{figure}

Figure 1 illustrates the bias, variance, and mean squared error of the harmonized estimator (see equations \eqref{simple_bias} and \eqref{simple_variance} in Proposition \ref{prop:simple}).
It also shows the  operating characteristics of other  estimators of $\theta_{1:K}$:
\begin{enumerate}[wide = 0.5\parindent, leftmargin = 0.5\parindent]
    \item[(a)] the \textit{RCT-only} estimator $\hat{\theta}^{(r)}_k = \bar{Y}_{k, 1}^{(r)} - \bar{Y}_{k, 0}^{(r)}$, for $k=1,\ldots,K$, which is unbiased, and has variance equal to $K \left( \frac{1}{n^{(r)}_{\cdot,0}} + \frac{1}{n^{(r)}_{\cdot,1}} \right) \phi^2 I_K$,
    
    \item[(b)] the \textit{pooled estimator} in equation \eqref{eq:basic_subgroup}, which has bias equal to $-q \gamma_{1:K}$ and variance equal to $K \frac{1}{n_{\cdot,0}^{(r)}} (2 - q) \phi^2 I_K$,
    
    \item[(c)] and an \textit{oracle estimator} $\hat{\theta}^{ora}_{1:K}$  (with known  $\mu_{1:K}$ values), 
    \begin{equation*}
        \hat{\theta}^{ora}_k = \bar{Y}_{k, 1}^{(r)} - \mu_k,
    \end{equation*}
    which is unbiased and has variance equal to $\frac{K}{n_{\cdot,0}^{(r)}} \phi^2 I_K$.
\end{enumerate}
In Figure \ref{fig:fig_1_new}, we examine three scenarios with different values of the subgroup-specific differences $\gamma_{1:k}$ between the EC and RCT control means: 
$\gamma_k = 0$ for each $k = 1, \dots, K$ (Scenario 1, no distortions), $\gamma_k = 1$ for each $k = 1, \dots, K$ (Scenario 2, SDM assumption holds), and $\gamma_{1:K} = (1 + \Delta_\gamma, 1 -\Delta_\gamma, 1+\Delta_\gamma, \dots)$ (Scenario 3, SDM assumption violated).
For the sample sizes we use $n^{(r)} = 100$ RCT patients and $n^{(e)} = 500$ EC patients, we have $K = 10$ subgroups, and   $\phi^2 = 1$.

Figure 1A shows how the bias of $\hat{\theta}^{h}_{1:K}$ converges to $0$ as $\lambda \to \infty$ in Scenario 2. The estimator $\hat{\theta}^{h}_{1:K}$ is suboptimal in settings where the bias terms $\gamma_{1:K}$ vary substantially across subgroups, such as Scenario 3. Recall that we introduced the harmonized estimator for inference on the treatment effects in small patient subgroups in disease settings where  heterogeneous bias mechanisms across subgroups (i.e., relevant variations of $\gamma_{1:k}$ across subgroups) appear implausible. 
We emphasize that in Figure 1A, Scenario 3  involves a high level of variability of $\gamma_k$ across subgroups to illustrate that when the SDM assumption is markedly violated harmonization produces biased estimates. Figure 1B shows that when the $\gamma_k$'s vary \textit{moderately} across subgroups ($\Delta_\gamma < 0.25$) the estimator $\hat{\theta}^h_{1:K}$ has lower root mean squared error (RMSE) than $\hat{\theta}^{(r)}$, despite  the  violation of the  SDM assumption. With $\Delta_\gamma < 0.25$ the violation of this assumption is less marked than in  Scenario 3, where the $\gamma_{k}$ parameters are highly heterogeneous ($\Delta_\gamma = 1$ in Panel A). The estimate $\hat{\theta}^{h}_{1:K}$ has  lower variance than $\hat{\theta}^{(r)}_{1:K}$.  Also, the variance of $\hat{\theta}^{h}_{1:K}$ is  similar to the variance of $\hat{\theta}^{(r+e)}_{1:K}$. 

We can  relax some of the assumptions in Proposition \ref{prop:simple}. First, the positive-definite matrix $\Sigma$ used in (\ref{harmonized_def}) to define the estimator $\hat{\theta}^h_{1:K}$ can be arbitrary. Second, the randomization ratio and the subgroup proportions in the RCT can be unbalanced, i.e. we consider $n_{\cdot, 1}^{(r)} \neq n_{\cdot, 0}^{(r)}$ and $\pi \neq (1/K, \dots, 1/K)$. We still assume that $\frac{n_{k,1}^{(r)}}{n_{k,0}^{(r)}}$ is constant across subgroups and $\frac{n_{k,\cdot}^{(r)}}{n^{(r)}}=\pi_k$, yet we allow the subgroup proportions in the EC data to  differ from those in the RCT; that is $\frac{n_{k,0}^{(e)}}{n^{(e)}}$ can be different from $\pi_k$. In this case, the harmonized estimator $\hat{\theta}^h_{1:K}$ remains normally distributed, and the bias and variance in expressions \eqref{simple_bias} and \eqref{simple_variance} become
\begin{equation} \label{relaxed_bias_variance} 
\begin{split}
    B^h &= - \left( I_K -  c \frac{\lambda}{ \lambda + c} \Sigma \pi \pi^\top \right) Q \gamma_{1:K}, \text{ and} \\
    V^h &= \phi^2  \left[ \frac{1}{n_{\cdot,1}^{(r)}} I_K - \frac{1}{n_{\cdot,0}^{(r)}} (I_K - Q) \right] \Pi^{-1} + \left( c \frac{\lambda}{ \lambda + c} \right)^2 \bar{q} \frac{1}{n_{\cdot, 0}^{(r)}} \phi^2 \Sigma \pi \pi^\top \Sigma
\end{split}
\end{equation}
respectively. Here $c = \left( \pi^\top \Sigma \pi \right)^{-1}$ as before, $Q$ is a diagonal $K \times K$ matrix with  $Q_{k,k} = \frac{n_{k,\cdot}^{(e)}}{n_{k,0}^{(r+e)}}$, $\bar{q} = \sum_{k=1}^K \pi_k Q_{k,k}$, 
and $\Pi$ is a diagonal $K \times K$ matrix with $\Pi_{k, k} = \pi_k$. When $\lambda = \infty$ and the SDM assumption holds with $\gamma_{1:K} = \gamma 1_{K \times 1}$ for some $\gamma \in \mathbb{R}$, the bias is $B^h = 0_{K \times 1}$ if we choose $\Sigma = \text{diag} \left( \frac{Q_{k,k}}{\pi_k} \right)$. This choice of $\lambda$ and $\Sigma$ is gives a BD harmonized estimator. Indeed, $\Sigma \pi $ becomes equal to  $Q 1_{K \times 1}$, and  $B^h $ can be rewritten as
$$ \left( I_K -  \left( \pi^\top \Sigma \pi \right)^{-1} \Sigma \pi \pi^\top \right) Q \gamma_{1:K} = \gamma \left( I_K - \left( \pi^\top \Sigma \pi \right)^{-1} \Sigma \pi \pi^\top \right) \Sigma \pi = 0_{K \times 1}. $$

For the $k$-th subgroup, the difference between the mean squared errors of the pooled estimator and the harmonized estimator is
\begin{equation}\label{mse_diff}
    MSE \left( \hat{\theta}_k^{(r+e)} \right) - MSE \left( \hat{\theta}_k^{h} \right) = q_k^2 \Big[ (\gamma_k - \bar{\gamma})^2 - \gamma_k^2 \Big] - \frac{ q_k^2 \phi^2}{n_{\cdot, 0}^{(r)} \left(\sum_{k=1}^K \pi_k q_k\right)},
\end{equation}
where $\bar{\gamma} = \frac{\sum_{k=1}^K \pi_k q_k \gamma_k}{\sum_{k=1}^K \pi_k q_k}$ is the weighted average of the subgroup-specific EC distortion parameters. The first term in the right hand of equation \eqref{mse_diff} is the difference between the squared biases of $\hat{\theta}^{(r+e)}_{1:K}$ and $\hat{\theta}^h_{1:K}$, and the second term is the difference between their variances. Pooling has a slightly lower variance compared to harmonization. In the absence of distortions (i.e., Scenario 1, with $\gamma_{k} = 0$ for all $k$)  both pooling and harmonization are unbiased. In contrast, in the presence of systematic distortions across subgroups (Scenario 2, $\gamma_{k} = \gamma \in \mathbb{R}$ for all $k$) pooling is biased while harmonization is unbiased.

\subsection{A quasi-Bayesian interpretation of the harmonized estimator} \label{subsection:cut}

We provide an interpretation of the harmonized estimator that builds on the modular Bayesian framework in \cite{jacob_better_2017} and \cite{ bayarri_modularization_2009}. In  this subsection we focus on a simple  model to exemplify the relationship between harmonization and Bayesian analyses. We consider two analysts: 

\noindent \textbf{Analyst 1} conducts a primary analysis and estimates $\theta$, the overall effect, using only the RCT data $D^{(r)}$ (without considering   subgroups). The analyst  uses the following Bayesian model:
\begin{equation} \label{analyst_1_model}
    \begin{alignedat}{2}
    \text{Likelihood:} & \hspace{2em}  Y_i^{(r)} | T_i^{(r)}, \mu, \theta  \overset{ind.}{\sim} N \left( \mu + \theta T_i^{(r)}, \sigma^2 \right), \\ 
    \text{Prior:} & \hspace{2em}  (\mu, \theta)  \sim N_2(m^{(A1)}, \tau^{(A1)}).
    \end{alignedat}
\end{equation}

To simplify the presentation the analyst chooses  some fixed value for  $\sigma^2$. 
The result of the analysis is the posterior  $p^{(A1)} \left( \theta | D^{(r)} \right) = p^{(A1)} \left( \theta | Y^{(r)}_{1:n^{(r)}}, T^{(r)}_{1:n^{(r)}} \right)$, conditional on the outcomes $Y^{(r)}_{1:n^{(r)}}$ and treatment assignments $T^{(r)}_{1:n^{(r)}}$ in the RCT. We use the superscript $(A1)$ to identify the posterior inference of Analyst 1. Note that $p^{(A1)} \left( \theta | D^{(r)} \right)$ does not involve the EC data.

\noindent \textbf{Analyst 2} conducts a subgroup analysis and estimates $\theta_{1:K}$ using the RCT and EC data $D^{(r+e)}$, thus augmenting the  subgroup-specific sample sizes of the RCT. The analyst uses the following Bayesian model:
\begin{equation} \label{analyst_2_model}
    \begin{alignedat}{2}
    \text{Likelihood:} & \hspace{2em}  Y_i^{(r)} | T_i^{(r)}, W_i^{(r)}, \mu_{1:K}, \theta_{1:K}  \overset{ind.}{\sim} N \left( \mu_{W_i^{(r)}} + \theta_{W_i^{(r)}} T_i^{(r)}, \phi^2 \right), \\
    & \hspace{2em} Y_i^{(e)} | W_i^{(e)}, \mu_{1:K}  \overset{ind.}{\sim} N \left( \mu_{W_i^{(e)}}, \phi^2 \right), \\
    \text{Prior:} & \hspace{2em}  (\mu_{1:K}, \theta_{1:K})  \sim N_{2K} \left( m^{(A2)}, \tau^{(A2)} \right). 
    \end{alignedat}
\end{equation}
This model assumes identical outcome distributions for the RCT controls and the EC population. Also in this case the  analyst chooses some fixed value for $\phi^2$. The analyst assumes that $\left( \nicefrac{n_{1,\cdot}^{(r)}}{n^{(r)}}, \dots, \nicefrac{n_{K,\cdot}^{(r)}}{n^{(r)}} \right)=\pi$. The posterior distribution is $p^{(A2)} \left(\theta_{1:K} | D^{(r)}, D^{(e)} \right)$. If the RCT controls and ECs have different outcome distributions then the model is misspecified. 

We can combine Analyst 1's inference on the overall effect $\theta$ and Analyst 2's inference on the subgroup-specific effects $\theta_{1:K}$ using the modular approach  discussed in \cite{jacob_better_2017}. Analyst 2 has a joint posterior on $ \theta_{1:K}$ and $\theta = \sum_{k=1}^K \pi_k \theta_k$, that can be factorized into the product of (i) the  posterior of $\theta$ and (ii) the  posterior of $\theta_{1:K}$ given $\theta$:
\begin{equation} \label{full_joint}
    p^{(A2)} \left(\theta, \theta_{1:K} | D^{(r)}, D^{(e)} \right) = p^{(A2)} \left(\theta | D^{(r)}, D^{(e)} \right) \times p^{(A2)} \left(\theta_{1:K} | \theta, D^{(r)}, D^{(e)} \right).
\end{equation}
Based on the modular framework we  obtain the following {\it cut} distribution (we use the term {\it cut} as in  \cite{jacob_better_2017}):
\begin{equation} \label{cut_def}
    p^{\text{cut}} \left( \theta_{1:K}, \theta | D^{(r)}, D^{(e)} \right) = 
    p^{(A1)} \left( \theta | D^{(r)} \right) \times
        p^{(A2)} \left( \theta_{1:K} | \theta, D^{(r)}, D^{(e)} \right) .
\end{equation}
This joint distribution is obtained by replacing $p^{(A2)} \left(\theta | D^{(r)}, D^{(e)} \right)$ in \eqref{full_joint} with $p^{(A1)} \left( \theta | D^{(r)} \right)$, which only uses the RCT data without subgroup information. The cut distribution of $\theta_{1:K}$ is based on both $D^{(r)}$ and $D^{(e)}$, and is coherent with the inference of the first analyst on $\theta$, which is based only on $D^{(r)}$. 

A simpler plug-in distribution can be obtained by ignoring Analyst 1's uncertainty about $\theta$, using ${\theta}^{(A1)} = E^{(A1)} \left( \theta | Y^{(r)}_{1:n^{(r)}}, T^{(r)}_{1:n^{(r)}} \right)$: 
\begin{equation}\label{plugcut_def}
    p^{\text{plug-in}} \left(\theta_{1:K} | D^{(r)}, D^{(e)} \right) = p^{(A2)} \left( \theta_{1:K} | \theta = {\theta}^{(A1)}, D^{(r)}, D^{(e)} \right).
\end{equation}

Both the cut distribution in \eqref{cut_def} and the plug-in distributions in \eqref{plugcut_def} produce point estimates of $\theta_{1:k}$ that coincide with the harmonized estimator. That is,
\begin{equation} \label{cut_equal}
     E^{\text{cut}} \left( \theta_{1:K} | D^{(r)}, D^{(e)} \right) = E^{\text{plug-in}} \left(\theta_{1:K} | D^{(r)}, D^{(e)} \right) = \hat{\theta}^h_{1:K},
\end{equation}
where $\hat{\theta}^h_{1:K}$ is obtained with inputs $\lambda = \infty$,  $\hat{\theta}^{(r)} = E^{(A1)} \left( \theta | D^{(r)} \right)$, $\hat{\theta}_{1:K}^{(r+e)} = E^{(A2)} \left( \theta_{1:K} | D^{(r)},D^{(e)} \right)$, and $\Sigma = Var^{(A2)} \left( \theta_{1:K} | D^{(r)},D^{(e)} \right)$ in the right-hand side of equation \eqref{harmonized_def}. Thus, the Bayesian view of our work suggests VD (variance-directed) harmonization, where $\lambda = \infty$ and $\Sigma$ measures the variability associated with $\hat{\theta}^{(r+e)}_{1:K}$. 

\begin{figure}[tb]
    \centering
    \includegraphics[width = \textwidth]{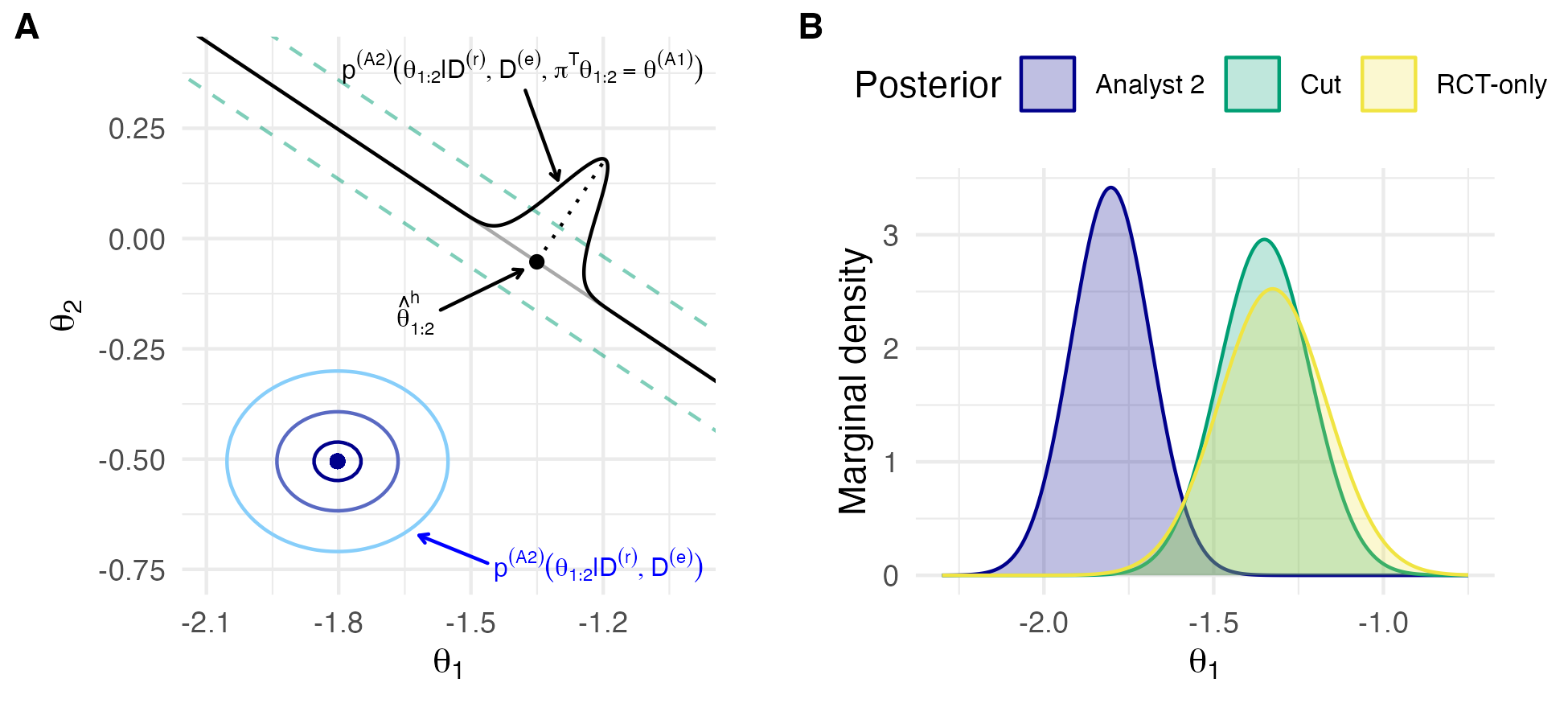}
    \caption{A quasi-Bayesian interpretation of the harmonized estimator with $K = 2$ subgroups. (A) Analyst 2's  posterior $p^{(A2)} \left( \theta_1, \theta_2 | D^{(r)}, D^{(e)} \right)$ in blue (mean  and  isolines). The plug-in posterior $p^{(A2)} \left( \theta_1, \theta_2 | D^{(r)}, D^{(e)}, \pi^\top \theta_{1:2} = {\theta}^{(A1)} \right)$ is the density in black, which has support equal to the linear subspace $\pi^\top \theta_{1:2} = {\theta}^{(A1)}$. The mean (black point) is equal to the harmonized estimator. The dashed lines show the linear subspaces $\pi^\top \theta_{1:2} = \theta$ with $\theta$ equal to the 25\% and 75\% quantiles of Analyst 1's posterior $p^{(A1)} ( \theta | D^{(r)} )$. 
    (B) The posterior density of $\theta_1$ under Analyst 2's joint model for RCT and EC data (blue), the cut distribution (equation \eqref{cut_def}, green), and a Bayesian subgroup analysis using only RCT data (yellow).}
    \label{cut_fig}
\end{figure}

The described relationship between the harmonized estimator and Bayesian analyses is illustrated in Figure \ref{cut_fig}A. In Figure 2B, we show the posterior density of $\theta_1$ under Analyst 2's joint model for the RCT and EC data (blue), the cut distribution (equation \eqref{cut_def}, green), and a Bayesian subgroup analysis using only RCT data (i.e. Analyst 2's  inference if $D^{(e)}$ is not included, yellow). We refer to the Supplementary Material for the details on the prior models and the simulation illustrated in this figure.

To verify equation \eqref{cut_equal}, it is sufficient to note that
\begin{align*}
    & \int E^{(A2)} \left( \theta_{1:K} | \pi^\top \theta_{1:K} = \theta, D^{(r)}, D^{(e)} \right) \cdot p^{(A1)} \left( \theta | D^{(r)} \right) d \theta \\
    & \hspace{2em} = E^{(A2)} \left( \theta_{1:K} | \pi^\top \theta_{1:K} = {\theta}^{(A1)}, D^{(r)}, D^{(e)} \right) \;\; \text{and},
\end{align*}
\begin{align*}
    E^{(A2)} \left( \theta_{1:K} | \pi^\top \theta_{1:K} = \hat{\theta}^{(A1)}, D^{(r)}, D^{(e)} \right) &= \underset{\pi^\top \theta_{1:K} = \hat{\theta}^{(A1)}}{\text{arg max}} p^{(A2)} \left( \theta_{1:K} | D^{(r)}, D^{(e)} \right) \\
    &= \underset{ \pi^\top \theta_{1:K} = \hat{\theta}^{(A1)}}{\text{arg min}}  \left( \theta_{1:K} - {\theta}^{(A2)}_{1:K} \right)^\top {\Sigma}^{-1}\left( \theta_{1:K} - {\theta}^{(A2)}_{1:K} \right),
\end{align*}
where ${\theta}_{1:K}^{(A2)} = E^{(A2)} \left( \theta_{1:K} | D^{(r)}, D^{(e)} \right)$ and ${\Sigma} = Var^{(A2)} \left( \theta_{1:K} | D^{(r)}, D^{(e)} \right)$. The first equality holds because $E^{(A2)} \left( \theta_{1:K} | \pi^\top \theta_{1:K} = \theta, D^{(r)}, D^{(e)} \right)$ is linear in $\theta$. The final two equalities hold because Analysts 1 and 2 use conjugate normal models. 
We note that the estimator in expression \eqref{cut_equal} is unbiased when $\gamma_1=\gamma_2\ldots= \gamma_{K}$ (SDM assumption) if the ratios $\frac{n_{k,0}^{(e)}}{n_{k,0}^{(r)}}$ and $\frac{n_{k,1}^{(r)}}{n_{k,0}^{(r)}}$ remain the same across subgroups $k=1,\ldots, K$ (see Supplementary Section S2.4.5).

Up to this point we have shown how the cut distribution and the harmonized estimator are connected when Analysts 1 and 2 have normal posterior distributions on $\theta$ and $\theta_{1:K}$. This discussion can be extended beyond the conjugate model that we considered. In particular, Analyst 2 may use a  Bayesian joint regression model for $D^{(r)}$ and $D^{(e)}$, for example a generalized linear model \citep{hobbs_commensurate_2012}, in which $\theta_{1:K}$ are subgroup specific treatment effects (i.e.,  the differences between the expected outcomes under experimental and control therapy 
 across the RCT subpopulations). 
 Analyst 1 may use another Bayesian model (with or without biomarker subgroups and covariate information) for $D^{(r)}$ alone, and $\theta$ is the average treatment effect in the overall population. If Bernstein-von Mises conditions \citep{ghosal_fundamentals_2017} are satisfied --- as in most   regression models --- then with sufficiently large $n^{(r)}$ and $n^{(e)}$ the relationships between posterior distributions and the harmonized estimators that we discussed (Figure \ref{cut_fig}A and equation \ref{cut_equal}) are preserved. 
          
\subsection{Interval estimation} \label{subsection:intervals}

We discuss  $(1 - \alpha)$-level  intervals  obtained through   harmonization.
We consider $        \hat{\theta}^{(r+e)}_k  = \bar{Y}_{k, 1}^{(r)} - \bar{Y}_{k, 0}^{(r+e)} 
$ and $        \hat{\theta}^{(r)} = \bar{Y}_{\cdot,1}^{(r)} - \bar{Y}_{\cdot,0}^{(r)}$ (equations \ref{eq:basic_subgroup} and \ref{eq:basic_overall}).
The confidence intervals of $\theta_k$ are computed assuming that $D^{(r)}$ and  $ D^{(e)}$ 
are distributed according to model \eqref{simple_model}. 
We describe three distinct approaches. 

\begin{enumerate}[(a)]
    \item  
    We can use the analytic  expression of the variance $V^h_{k,k}$ (equation \ref{relaxed_bias_variance}), and compute 
    \begin{equation} \label{harm_int}
        \left( \hat{\theta}^h_k - z_{1 - \nicefrac{\alpha}{2}} \sqrt{V^{h}_{k,k}}, \hspace{0.75em} \hat{\theta}^h_k + z_{1 - \nicefrac{\alpha}{2}} \sqrt{V^h_{k,k}} \right),
    \end{equation}
    where $z_{1-\nicefrac{\alpha}{2}}$ is the $\left( 1 - \nicefrac{\alpha}{2} \right) \times 100 \%$ quantile of the standard normal distribution. Here 
    we are assuming implicitly that $\hat{\theta}^h_k$ is an unbiased estimate (see subsection \ref{subsection:simple_bv}). Standard central limit arguments imply that the confidence interval can still be considered in settings where the 
    assumption of normally distributed outcomes in model \eqref{simple_model} is violated.
    
    \item Based on the connection between the harmonized estimator and Bayesian analyses illustrated in subsection \ref{subsection:cut}, we can use the cut distribution in \eqref{cut_def} to compute the interval
    \begin{equation} \label{cut_int}
        \left( \hat{\theta}^h_k - z_{1 - \alpha/2} \sqrt{V^{(cut)}_{k,k}}, \hspace{0.75em} \hat{\theta}^h_k + z_{1 - \alpha/2} \sqrt{V^{(cut)}_{k,k}} \right),
    \end{equation}
    where $V_{k,k}^{(cut)}$ is the variance of $\theta_k$ under the cut distribution. If Analysts 1 and 2 use the conjugate models described in subsection \ref{subsection:cut}, then the
     covariance matrix is 
    \begin{equation*}
        V^{(cut)} = \Sigma^{(A2)}_{1:K} + \left( \frac{1}{\Sigma^{(A2)}_\theta} \right)^2 \left( \Sigma^{(A1)}_\theta - \Sigma^{(A2)}_\theta \right) \Sigma_{1:K}^{(A2)} \pi \pi^\top \Sigma_{1:K}^{(A2)},
    \end{equation*}
    where $\Sigma_{1:K}^{(A2)}$ is the posterior covariance matrix of $\theta_{1:K}$ under Analyst 2's model, $\Sigma_\theta^{(A2)}$ is the posterior variance of $\theta$ under Analyst 2's model, and $\Sigma_\theta^{(A1)}$ is the posterior variance of $\theta$ under Analyst 1's model. Moreover, $V^{(cut)} = V^h$ under the following conditions: 
    1) Analysts 1 and 2 use flat priors ($\tau^{(A1)} $ and $\tau^{(A2)}$ are diagonal with diverging diagonal entries), 
    2) the variance parameters $\sigma^2$ and $\phi^2$ in models \eqref{analyst_1_model}, \eqref{analyst_2_model}, and \eqref{simple_model} (underlying $V^h$) are identical, 
    and 3) $\lambda = \infty$ and  $\Sigma$ is diagonal with entries proportional to $\frac{1}{n_{k, 1}^{(r)}} + \frac{1}{n_{k, 0}^{(r+e)}}$. 
    Thus the intervals in expressions \eqref{cut_int} and \eqref{harm_int} become identical.

    \item  The bootstrap method 
    \citep{efron_bootstrap_1986} can also be used to construct confidence intervals. We can generate $a = 1, \dots, R$ replicates $\left( D^{(r)}_a, D^{(e)}_a \right)$ through a parametric bootstrap  (using model \ref{simple_model}). 
    For each  replicate, the harmonized estimator $\hat{\theta}_{1:K, a}^{h}$ is computed. 
    This allows us to estimate the variability of $\hat{\theta}_{k}^{h}$ through the empirical distribution $\hat{G}_k$ of the $\hat{\theta}_{k, a}^{h}$ replicates. 
    We can report a confidence interval for $\theta_k$ that is centered at  $\hat\theta_k^h$ and has width that matches the distance between the quantiles $\hat{G}^{-1}_k \left( \nicefrac{\alpha}{2} \right)$ and $\hat{G}^{-1}_k \left(1- \nicefrac{\alpha}{2} \right)$.     
\end{enumerate}

\begin{figure}[tbp!]
    \centering
    \includegraphics[width = 0.95\textwidth]{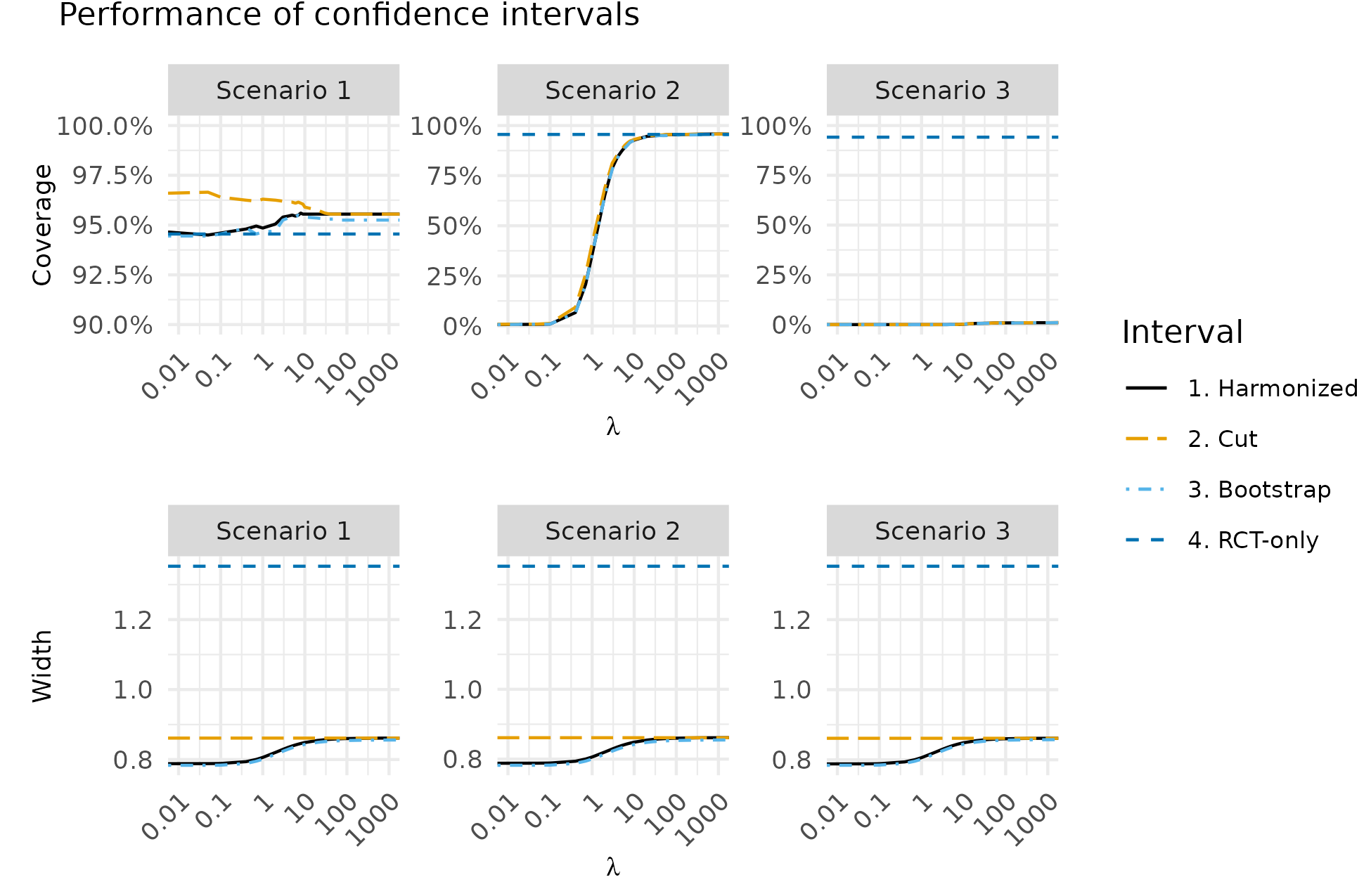}
    \caption{Coverage and width of 95\% confidence intervals. We use the procedures described in subsection \ref{subsection:intervals} to compute the intervals. We consider the same scenarios as in Figure \ref{fig:fig_1_new}. Coverage and average width are computed using 2,000 Monte Carlo replicates for each scenario. On the x-axis  we vary the $\lambda$ value that defines the harmonized estimator.}
    \label{fig:int_fig}
\end{figure}

In Figure \ref{fig:int_fig} we compare the coverage and average length of these intervals through simulations. We also consider the usual confidence interval of $\theta_k$ based  only on  RCT data:
\begin{equation*} \label{rct_int}
    \left(\hat{\theta}^{(r)}_{k} - z_{1 - \alpha/2} 
\times   \sqrt{ \text{Var} \left(\hat{\theta}^{(r)}_{k} \right) }, \hspace{0.75em} \hat{\theta}^{(r)}_{k} + z_{1 - \alpha/2} 
\times   \sqrt{ \text{Var} \left(\hat{\theta}^{(r)}_{k} \right) } \right),
\end{equation*}
where $\hat{\theta}^{(r)}_k = \bar{Y}_{k, 1}^{(r)} - \bar{Y}_{k, 0}^{(r)}$. We include this interval for comparison.
We simulate data using the same scenarios used for Figure \ref{fig:fig_1_new}. 
For the harmonized estimator we use $\Sigma = diag \left( \frac{Q_{k,k}}{\pi_k} \right)$, 
and $\left( \hat\theta^{(r)},  \hat\theta^{(r+e)}_{1:K} \right)$ as defined by expressions \eqref{eq:basic_overall} and \eqref{eq:basic_subgroup}. 
To compute the cut distribution  we set $\tau^{(A1)} = \text{diag}(10^4)$ and $\tau^{(A2)} = \text{diag}(10^4)$. For each scenario we simulated 2,000 trials. 
We illustrate results for subgroup  $k = 1$. Figure \ref{fig:int_fig} shows that in Scenarios 1 and 2, in which  $\hat{\theta}^h_{1}$ is unbiased for  $\lambda=\infty$,  the intervals  have coverage  nearly identical to the nominal $1-\alpha$ level (when $\lambda>10$). 
In contrast in Scenario 3, where the SDM assumption is markedly violated and $\hat{\theta}^h_1$ is biased, as expected the coverage is far below the nominal value (close to 0 \%).
The harmonized and cut intervals centered on $\hat{\theta}^h_{1}$ are substantially shorter than the intervals that we compute using only RCT data.
The poor coverage in Scenario 3 emphasizes that harmonization is not sufficient to remove the bias of $\hat{\theta}^{(r+e)}_{1:K}$ when the SDM assumption (i.e., $\gamma_1 = \dots = \gamma_K$, equation \eqref{systematic_distortion_sec_2}) is violated. Importantly, data analyses can detect violations of the SDM assumption (see for example Supplementary Section S3.8). Figure \ref{fig:int_fig} confirms that the analyst should consider harmonization if (i)  high quality external control data (see \cite{corrigan-curay_real-world_2018} for explicit criteria) are available, (ii) the violation of the SDM assumption based on context-specific considerations is implausible, and (iii) the datasets do not provide evidence of a violation of the SDM assumption.

When an analytic expression of the interval, based on exact or asymptotic arguments (e.g., equation \eqref{harm_int}), is available, the computing time is lower than bootstrapping. However, if the  distribution of  $\hat{\theta}^{h}_{1:K}$  is not available, then the bootstrap interval is an easy-to-interpret and practical alternative. In Figure \ref{fig:int_fig}, we observe negligible differences between the approaches that we considered to provide confidence intervals. 

\section{Harmonized estimators and regression models}

In this section we  discuss harmonized estimators  that allow the investigator  to account for pre-treatment covariates $X$.
  The distributions of relevant prognostic variables, such as patient age, may be different in the RCT and  EC populations. 
Recall that the datasets are $D^{(r)} = \left( Y_i^{(r)}, T_i^{(r)}, W_i^{(r)}, X_i^{(r)} \right)_{i\le n^{(r)}}$ for the RCT and $D^{(e)} = \left(Y_i^{(e)}, T_i^{(e)}, W_i^{(e)}, X_i^{(e)} \right)_{i\le n^{(e)}}$ for the EC.
Here  $X$ includes $d$ pre-treatment covariates, such as demographics. 
Also, the harmonized estimator $\hat{\theta}_{1:K}^h$ is  a transformation of  (i) $\hat{\theta}_{1:K}^{(r+e)}$, an  estimator of $\theta_{1:K}$ (the subgroup-specific treatment effects) based on  $D^{(r)}$ and $D^{(e)}$, and (ii) $\hat{\theta}^{(r)}$, an estimator of $\theta = \pi^\top \theta_{1:K}$ (the average treatment effect in the randomized population)
based  only on $D^{(r)}$.
The examples in this Section include linear regression, logistic regression, and propensity score weighted logistic regression. In each case we show how to obtain a BD harmonized estimator that is approximately unbiased when the SDM assumption holds. These results are unified in one proposition (see Proposition S2) on weighted generalized linear models (GLMs) in Section S3.6 of the Supplementary Material. The analytic expressions of $\Sigma$ for BD harmonization in this Section are special cases of Propositions S3 and  S5 in the Supplementary Material.

\subsection{Linear model} \label{subsection:linear_model}
We first discuss the harmonized estimator in a simple setting, where the data are analyzed through  linear models. In particular, we consider $\hat{\theta}_{1:K}^h$ defined by equation \eqref{harmonized_def} with the following inputs. 
\vspace{-0.5em}
\begin{enumerate}
    \item[(a)] The estimator $\hat{\theta}^{(r)}$ is an unbiased estimator, like the simple difference of means in equation \eqref{eq:basic_overall}. In the remainder of this subsection we consider $\hat{\theta}^{(r)}$ based on ordinary least squares (OLS), assuming that 
    \begin{equation} \label{linear_model_overall_ols}
        E \left(Y_i^{(r)} | T_i^{(r)}, X_i^{(r)} \right) = \mu + \theta T_i^{(r)} + \beta^\top X_i^{(r)}.
    \end{equation}
    Sufficient conditions   for   $\hat{\theta}^{(r)}$ to be an
    unbiased estimator   of $E(Y_i^{(r)}\mid T_i^{(r)}=1)-E(Y_i^{(r)}\mid T_i^{(r)}=0)$
    are provided in the Supplementary Material.
    \item[(b)] The estimator $\hat{\theta}^{(r+e)}_{1:K}$ is computed with OLS, assuming
    \begin{equation} \label{linear_model_ols}
        \begin{split}
            E \left(Y_i^{(r)} | W_i^{(r)}, T_i^{(r)}, X_i^{(r)} \right) &= \mu_{W_i^{(r)}} + \theta_{W_i^{(r)}} T_i^{(r)} + \beta^\top X_i^{(r)},\\
            E \left(Y_i^{(e)} | W_i^{(e)}, X_i^{(e)}\right) &=  \mu_{W_i^{(e)}} + \beta^\top X_i^{(e)}, \;\;\text{and}\\
            Var \left(Y_i^{(r)} | W_i^{(r)}, T_i^{(r)}, X_i^{(r)} \right) &= Var\left(Y_i^{(e)} | W_i^{(e)}, X_i^{(e)}\right) = \phi^2.
        \end{split}
    \end{equation}
    In this model, conditional on the covariates $X$, the outcomes 
    in the external and RCT control groups have identical mean and variance.
\end{enumerate}
We evaluate the resulting  harmonized estimator when the assumptions of model \eqref{linear_model_ols} are violated. In particular we consider a   \textit{true} data-generating model 
with conditional moments
\begin{equation}
\label{linear_model}
    \begin{split}
        E \left(Y_i^{(r)} | W_i^{(r)}, T_i^{(r)}, X_i^{(r)} \right) &= \mu_{W_i^{(r)}} + \theta_{W_i^{(r)}} T_i^{(r)} + \beta^\top X_i^{(r)},\\
        E \left(Y_i^{(e)} | W_i^{(e)}, X_i^{(e)}\right) &=  \mu_{W_i^{(e)}} + \gamma_{W_i^{(e)}} + \beta^\top X_i^{(e)}, \;\;\text{and} \\
        Var \left(Y_i^{(r)} | W_i^{(r)}, T_i^{(r)}, X_i^{(r)} \right) &= Var\left(Y_i^{(e)} | W_i^{(e)}, X_i^{(e)}\right) = \phi^2.
    \end{split}
\end{equation}
This  conditional distribution differs from   model  \eqref{linear_model_ols} and the parameters $\gamma_{1:K}$  allow the 
external and RCT controls  to have different conditional  means. 
Here the SDM assumption  is formalized by  the following equality 
\begin{equation} \label{systematic_distortion_sec_3_1}
    \gamma_k = E \left( Y_i^{(e)} | W_i^{(e)} = k, X_i^{(e)} = x \right) - E \left( Y_i^{(r)} | W_i^{(r)} = k, X_i^{(r)} = x, T_i^{(r)} = 0 \right) = \gamma, \
\end{equation}
for all subgroups $k = 1, \dots, K$ and some $\gamma \in \mathbb{R}$.
 
We evaluate the harmonized estimators  conditioning on  individual covariates $X$  and treatment assignment. 
It can be shown under minimal design assumptions that $\hat{\theta}^{(r)}$ is an unbiased estimator of $\theta = \pi^\top \theta_{1:K}$, and that
\begin{equation} \label{linear_r_e_bias}
\begin{split}
    & Bias \left( \hat{\theta}_{1:K}^{(r+e)}, \theta_{1:K} \right) = B \gamma_{1:K}, \hspace{2em} \text{with} \\
    & B = \bmat{0_{K \times K}, & I_K, & 0_{K \times d}} \left( M_1^\top M_1 \right)^{-1} M_1^\top M_2,
\end{split}
\end{equation}
where $\bmat{0_{K \times K}, & I_K, & 0_{K \times d}}$ is a block matrix of  dimension $K \times (2K + d)$, $I_K$ is the $K \times K$ identity matrix, and $(M_1 , M_2)$ form design matrices. In particular $M_1$ is the $n^{(r+e)} \times (2K + d)$ design matrix for model \eqref{linear_model_ols}; it contains columns corresponding to $\mu_{1:K}, \theta_{1:K}$, and $ \beta$. The full design matrix for model \eqref{linear_model} is $\bmat{M_1, & M_2}$; 
$M_2$ contains the columns corresponding to $\gamma_{1:K}$ and has dimension $n^{(r+e)} \times K$.
Moreover,
\begin{align} \label{linear_r_e_cov}
    Var \left( \hat{\theta}^{(r+e)}_{1:K} \right) &= \phi^2 \bmat{0_{K \times K}, & I_K, & 0_{K \times d}} \left(M_1^\top M_1 \right)^{-1} \bmat{0_{K \times K}, & I_K, & 0_{K \times d}}^\top, \notag \\
    Var \left( \hat{\theta}^{(r)} \right) &= \phi^2 \left(M_0^\top M_0 \right)^{-1}_{2,2},  \\
    \hspace{-1em} Cov \left( \hat{\theta}^{(r+e)}_{1:K}, \hat{\theta}^{(r)} \right) &=  
    \\ 
     %
    \phi^2 \bmat{0_{K \times K}, & I_K, & 0_{K \times d}} & \left(M_1^\top M_1 \right)^{-1} M_1^\top \bmat{I_{n^{(r)}}, & 0_{n^{(e)} \times n^{(r)}}}^\top M_0 \left(M_0^\top M_0 \right)^{-1} \bmat{0, & 1, & 0_{1 \times d}} ^\top, \notag
\end{align} 
where $M_0$ is the $n^{(r)} \times (2 + d)$ design matrix for model \eqref{linear_model_overall_ols}.

We use these equations and the next three propositions to discuss $\hat{\theta}^h_{1:K}$.  The propositions can be useful to examine harmonized estimators in settings beyond the linear model, as we will see in later subsections. We  use the notation $S = Var \left( \hat{\theta}^{(r+e)}_{1:K} , \hat{\theta}^{(r)} \right)$
and $u = c \frac{\lambda}{\lambda + c} \Sigma \pi$. The harmonized estimator in equation \eqref{harmonized_simple} can  be rewritten as
$$ \hat{\theta}_{1:K}^{h} = \hat{\theta}_{1:K}^{(r+e)} + \left( \hat{\theta}^{(r)} - \pi^\top \hat{\theta}^{(r+e)}_{1:K} \right) u.$$

\begin{proposition} \label{prop:rid_1}
   If the following two assumptions 
    \begin{enumerate}[wide = 0.5\parindent, leftmargin = 0.5\parindent, label = (A\arabic*)]
        \item \label{assumption:r_unbiased} \hspace{0.1em} $E \left( \hat{\theta}^{(r)} \right) - \pi^\top \theta_{1:K} = 0$, and 
        \item \label{assumption:re_bias_linear} \hspace{0.1em} $E \left( \hat{\theta}^{(r+e)}_{1:K} \right) - \theta_{1:K} = B \gamma_{1:K}$, where $B$ is a $K \times K$ matrix, 
    \end{enumerate}
    hold,
    then
    \begin{equation} \label{rid_bias}
        Bias \left( \hat{\theta}_{1:K}^h, \theta_{1:K} \right) = \left( I_K - u \pi^\top \right) B \gamma_{1:K}
    \end{equation}
    and 
    \begin{equation} \label{rid_variance}
        Var \left( \hat{\theta}_{1:K}^h \right) = P S P^\top,
    \end{equation}
    where $P = \bmat{I_K - u \pi^\top, & u}$.
\end{proposition}

For the  setting that we are discussing,  with  $\hat{\theta}_{1:K}^{(r+e)}$ and $\hat{\theta}^{(r)}$ based on models \eqref{linear_model_ols} and \eqref{linear_model_overall_ols} respectively, and the data generating model summarized by expression  \eqref{linear_model}, this proposition allows us to derive the bias and variance of $\hat{\theta}^h_{1:K}$. Assumption \ref{assumption:r_unbiased} holds under minimal design assumptions and  \ref{assumption:re_bias_linear} is concordant with equation \eqref{linear_r_e_bias}. Thus we only need to plug the values of $\phi$, $M_0$, $M_1$, $M_2$, and $\gamma_{1:k}$ in to equations (\ref{linear_r_e_bias}-\ref{rid_variance}). Moreover, analytic expressions  of the conditional  bias and variance, analogous to equations \eqref{rid_bias} and \eqref{rid_variance}, can be easily obtained when $ \hat{\theta}^{(r)} = \bar{Y}_{\cdot,1}^{(r)} - \bar{Y}_{\cdot,0}^{(r)}$ and/or for parameterizations of the data generating model \eqref{linear_model} and design matrices that violate 
the assumption  \ref{assumption:r_unbiased}, because the harmonized estimator remains a linear transformation of the outcome data 
$Y_i^{(r)}$ and $Y_i^{(e)}$. 
The same comment holds if $\hat{\theta}^{(r)}=\pi^{T}  \hat{\theta}^{(r)}_{1:K}$, with group-specific estimates 
of the treatment effects $ \hat{\theta}^{(r)}_{1:K} $ obtained through OLS by restricting model \eqref{linear_model_ols} only to the RCT data.

The next result indicates for which fixed choice of $\lambda$ and $\Sigma$ the harmonized estimator  is unbiased under the SDM assumption (i.e., the differences between the EC and  RCT control groups are constant
across subpopulations). We use the notation $b = (b_1, \dots, b_K) = B 1_{K \times 1}$. 
\begin{proposition} \label{prop:rid_2}
If assumptions \ref{assumption:r_unbiased} and \ref{assumption:re_bias_linear} are satisfied, $\gamma_{1:K} = \gamma 1_{K \times 1}$ for some $\gamma \in \mathbb{R}$, and $b \neq 0_{K \times 1}$, then $\hat{\theta}^h_{1:K}$ is unbiased (i.e. $E \left( \hat{\theta}^h_{1:K} \right) = \theta_{1:K}$) if and only if $\lambda = \infty$ and $\Sigma \pi = \kappa b$ for some $\kappa \neq 0 $.
\end{proposition}
This proposition can be used to construct a BD harmonized estimator that is unbiased under the SDM assumption \eqref{systematic_distortion_sec_3_1}. The main requirement is to solve the equation $\Sigma \pi = \kappa b$. We only need the design matrices and it is not necessary to know the regression coefficients of model \eqref{linear_model}. For example, if the signs of the elements of $b = (b_1, \dots, b_K)$ are all the same, then taking $\Sigma = \text{diag} \left( \frac{|b_k|}{\pi_k} \right)$ solves the equation. Recall that $b$
is proportional to the bias of $\hat{\theta}_{1:K}^{(r+e)}$ when $\gamma_{1:K} \propto 1_{K \times 1}$; therefore one can  expect the individual entries $b_k$ to have identical signs. 
The next proposition indicates that we can choose $\Sigma$ that makes $\hat{\theta}_{1:K}^h$ unbiased when $\gamma_{1:K} \propto 1_{K \times 1}$. 
\begin{proposition}
\label{prop:sigma_existence}
There exists a positive definite matrix $\Sigma \in \mathbb{R}^{K\times K}$ such that $\Sigma \pi = \kappa b$ for some $\kappa \neq 0 $ if and only if $\pi^\top b \neq 0$.
\end{proposition}
If many positive matrices $\Sigma \in \mathbb{R}^{K \times K}$ solve the equation $\Sigma \pi = \kappa b$, with $\lambda = \infty$, they all define the same harmonized estimator. In other words, the BD harmonized estimator is unique. Indeed, when $\lambda = \infty$, the estimator $\hat{\theta}^h_{1:K}$  depends on $\Sigma$  only through  $\left( \pi^\top \Sigma \pi \right)^{-1} \Sigma \pi$ (see equation \ref{harmonized_simple}), which becomes $\left( \pi^\top b \right)^{-1} b$ if $\Sigma \pi = \kappa b$. In other words, the harmonized estimator   that satisfies the equation  $E \left( \hat{\theta}^h_{1:K} \right) = \theta_{1:K}$ is unique. 
We also note that the 
condition $\pi^\top b \neq 0$ in Proposition \ref{prop:sigma_existence} is not particularly restrictive. For example, if the covariates $X_i^{(r)}$ and $X_i^{(e)}$ have a continuous distribution, then
the event $\pi^\top b \neq 0$  has probability 1. 
 
Figure 4 illustrates  some simulations. 
We simulated data from the linear model \eqref{linear_model}, with $\gamma_{1:K}$ defined  as in  Scenarios 1-3  in Figure \ref{fig:fig_1_new}. 
We generated a  single  covariate (i.e. $d = 1$),  $X_i^{(r)} \overset{iid}{\sim} N(0, 1)$ for $i = 1, \dots, n^{(r)}$ in the RCT and $X_i^{(e)} \overset{iid}{\sim} N(2, 1)$ for $i = 1, \dots, n^{(e)}$ in the EC.   The vector of covariate values was the same across simulation replicates. 
The regression coefficient is $\beta = 0.5$. 
Beyond  the addition of  the covariate $X_i$  all simulation details (RCT design, EC design, and treatment effects) are the same as in Figure \ref{fig:fig_1_new}. 
We compare three estimators: 
(i) the pooled OLS estimator $\hat{\theta}^{(r+e)}_{1:K}$ based on  model \eqref{linear_model_ols}, 
(ii) the BD harmonized estimator that we described, using $\lambda = \infty$ and $\Sigma = \text{diag} \left( \nicefrac{|b_k|}{\pi_k} \right)$ (``BD harmonized''), 
and (iii) the RCT-only OLS estimator $\hat{\theta}^{(r)}_{1:K}$ based on the subgroup-specific model \eqref{linear_model_ols} excluding the EC data. 
The harmonized estimator has variance comparable to the pooled estimator and it is more accurate than the RCT-only analysis in Scenarios 1 and 2. In Scenario 3, with $\gamma_{1:K} \not \propto 1_{K \times 1}$, as expected the harmonized estimator inherits the bias of the pooled estimator, although  moderately reduced.

See Supplementary Figure S3 (Section S2.8) for similar comparisons of these estimators that focus on the bias, variance, and RMSE when the trial size $n^{(r)} = 100$ is fixed (with $n^{(r)}_{.,0} = 33$  and  $n^{(r)}_{.,1} = 67$), the EC sample size $n^{(e)}$ increases from $33$ to $627$ (i.e., $q = n^{(e)}_{.,0}(n^{(e)}_{.,0} + n^{(r)}_{.,0})^{-1}$ increases from 0.5 to 0.95) and the SDM assumption is violated. The bias and RMSEs of both the pooled and harmonized estimators increase with $n^{(e)} $. As expected this trend is less marked for the harmonized estimator compared to the pooled estimator. Importantly, when $n^{(e)}$ is large other definitions of $\hat{\theta}^{(r+e)}_{1:K}$, beyond pooling (e.g., Bayesian estimators based on hierarchical borrowing or power priors), can be used to reduce the outlined risks associated with distortion mechanisms that violate the SDM assumption. 

\begin{figure}[tbp!]
    \centering
    \includegraphics[width = 0.95\textwidth]{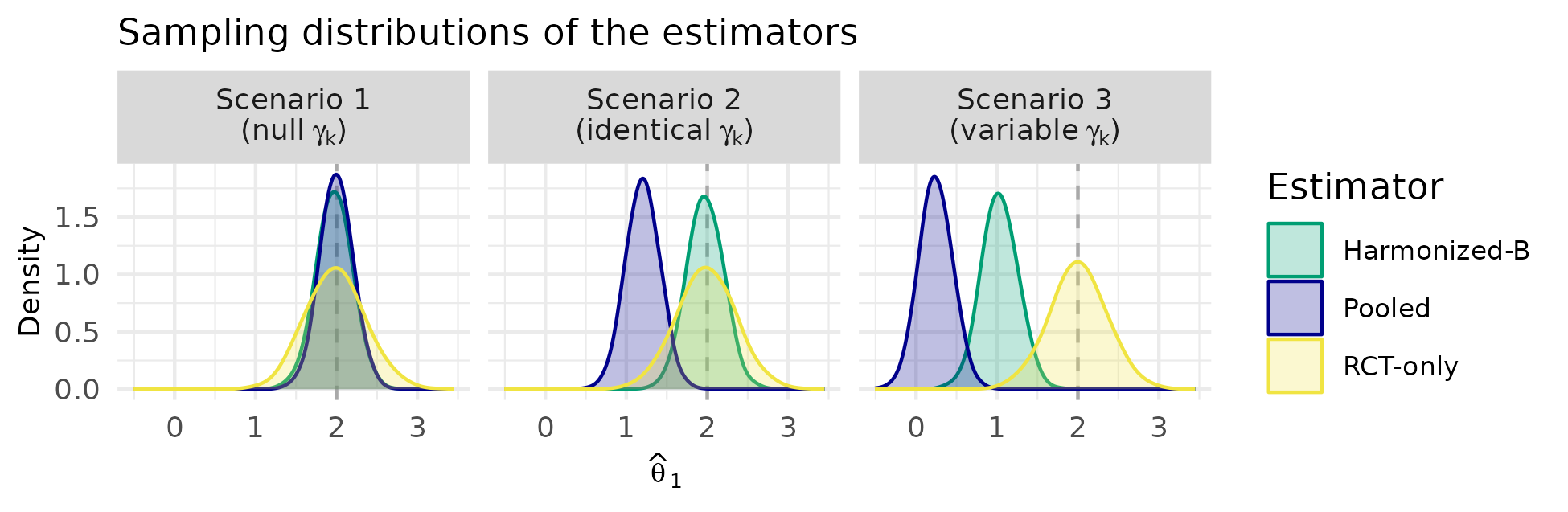}
    \caption{Sampling distributions of three estimators of $\theta_1$. The densities shown are kernel  estimates based on 2,000 simulated trials per  scenario. The dashed lines indicate the true value of  $\theta_1$. The scenarios present different configurations  of $\gamma_{1:K}$, the subgroup-specific differences between the EC and RCT control mean outcomes conditional on $X$.  These configurations are identical to those in Figure \ref{fig:fig_1_new}. Other than the addition of $X_i$  all simulation details are the same as in Figure \ref{fig:fig_1_new}.}
    \label{fig:linear_density}
\end{figure}

In these simulations we also consider the {\it cut } estimator (equation \eqref{cut_def} in subsection \ref{subsection:cut}). Here Analyst 1 uses model \eqref{linear_model_overall_ols} and flat priors for $\mu$, $\theta$, and $\beta$, and Analyst 2 uses model \eqref{linear_model_ols} with flat priors for $\mu_{1:K}$, $\theta_{1:K}$, and $\beta$. As before, the mean of the resulting {\it cut distribution} is a VD harmonized estimator with $\lambda = \infty$ and $\Sigma$ equal to Analyst 2's posterior variance of $\theta_{1:K}$. With flat priors $\hat{\theta}^{(r)} = E^{(A1)} \left( \theta | D^{(r)} \right)$ and $\hat{\theta}_{1:K}^{(r+e)} = E^{(A2)} \left( \theta_{1:K} | D^{(r)},D^{(e)} \right)$ are identical to the OLS estimates discussed throughout this subsection. In our simulations the cut estimates are nearly identical ($R^2 > 0.99$, see Supplementary Material) to the BD harmonized 
estimates in Figure \ref{fig:linear_density}. 

\subsection{Logistic regression} \label{subsection:logistic_model}
We  now consider harmonized estimators for binary outcomes.  The estimator   $\hat{\theta}^{(r+e)}_{1:K}$  is based on the \textit{pooled} logistic regression model: 
\begin{equation}
    \begin{split} \label{model:logistic_mle}
        p \left( Y^{(r)}_i = 1 | W^{(r)}_i, T^{(r)}_i, X^{(r)}_i \right) &= g\left( \nu_{W^{(r)}_i} + \eta_{W^{(r)}_i}T^{(r)}_i + \beta^\top X^{(r)}_i \right),\\
        p \left( Y^{(e)}_i = 1 | W^{(e)}_i, X^{(e)}_i \right) &= g\left( \nu_{W^{(e)}_i} + \beta^\top X^{(e)}_i \right),
    \end{split} 
\end{equation}
where $g(x) = \frac{1}{1 + e^{-x}}$,  and the outcomes are independent. 
The maximum likelihood estimates (MLEs)  are   $\left( 
\hat{\nu}_{1:K}^{(r+e)}, \hat{\eta}_{1:K}^{(r+e)}, \hat{\beta}^{(r+e)} \right)$. 
We consider the treatment effects
\begin{align} \label{logistic_theta_k}
    \theta_k &= E \left( Y_i^{(r)} | T_i^{(r)} = 1, W_i^{(r)} = k \right) - E \left( Y_i^{(r)} | T_i^{(r)} = 0, W_i^{(r)} = k \right),
\end{align}
and the estimates  
\begin{equation} \label{logistic_theta_k_re}
    \hat{\theta}_k^{(r+e)} = \frac{1}{n^{(r)}_{k,\cdot}} \sum_{i : W_i^{(r)} = k} \left[ g \left(\hat{\nu}^{(r+e)}_k + \hat{\eta}^{(r+e)}_k + \hat{\beta}^{(r+e)^\top} X_i^{(r)} \right) - g \left( \hat{\nu}^{(r+e)}_k + \hat{\beta}^{(r+e)^\top} X_i^{(r)} \right) \right].
\end{equation}
We  also specify a consistent estimator $\hat{\theta}^{(r)}$ of $\theta = \pi^\top \theta_{1:K}$. 
Then, we examine the resulting harmonized estimator (equation \ref{harmonized_simple}) when 
the individual records in $D^{(r)}$ and $D^{(e)}$ are independent replicates, with
%
\begin{equation}
    \begin{split} \label{model:logistic_true}
    p \left( Y^{(r)}_i = 1 | W^{(r)}_i, T^{(r)}_i, X^{(r)}_i \right) & = g\left( \nu_{W^{(r)}_i} + \eta_{W^{(r)}_i}T^{(r)}_i + \beta^\top X^{(r)}_i \right), \;\text{ and}\\
    p \left( Y^{(e)}_i = 1 | W^{(e)}_i, X^{(e)}_i \right) & =  g\left( \nu_{W^{(e)}_i} + \delta_{W^{(e)}_i} + \beta^\top X^{(e)}_i \right).
    \end{split}
\end{equation}

Expression \eqref{model:logistic_true} summarizes the true outcome distributions and expression \eqref{model:logistic_mle} describes the working model used to compute $\hat{\theta}^{(r+e)}_{1:K}$.  
If the parameters $\delta_k$ are different from zero, then the RCT and EC controls  have different subgroup-specific response rates after conditioning on the covariates $X$. 
Here the SDM assumption  is  formalized by  the  equality
\begin{equation} \label{systematic_distortion_sec_3_2}
    \delta_k = g^{-1} \left( E \left( Y_i^{(e)}  |  X_i^{(e)} = x, W_i^{(e)} = k \right) \right) - g^{-1} \left( E \left( Y_i^{(r)}  |  X_i^{(r)} = x, W_i^{(r)} = k, T_i^{(r)} = 0 \right) \right) = \delta, 
\end{equation}
for all subgroups $k = 1, \dots, K$ and some $\delta \in \mathbb{R}$.

We provide results similar to those  discussed for the linear model in subsection \ref{subsection:linear_model}. Here we consider  the asymptotic properties of a sequence of harmonized estimates,  when the sample sizes increase:
\vspace{-0.5em}
\begin{enumerate}[wide = 0.5\parindent, leftmargin = 0.5\parindent, label = (B\arabic*)]
      \item \label{assumption:logistic_n}  Both $ n^{(r)} \to \infty$ and $n^{(e)} \to \infty$.  The ratio $\frac{n^{(e)}}{n^{(r+e)}}$ converges to a value in $(0, 1)$. Also, $\frac{n^{(r)}_{k, 1}}{n^{(r)}_{\cdot, 1}}$, $\frac{n^{(r)}_{k, 0}}{n^{(r)}_{\cdot, 0}}$, and $\frac{n^{(e)}_{k, 0}}{n^{(e)}}$  converge to values in $(0, 1)$ for each $k = 1, \dots, K$.    
      
      \item \label{assumption:logistic_convergence}  
      We also assume that $\left( \hat{\nu}_{1:K}^{(r+e)}, \hat{\eta}_{1:K}^{(r+e)}, \hat{\beta}^{(r+e)}, \hat{\theta}_{1:K}^{(r+e)} \right)$ converges in probability to some point $\left( \nu_{1:K}^\circ, \eta_{1:K}^\circ, \beta^\circ, \theta_{1:K}^\circ \right)$, which might differ from the true $\left( \nu_{1:K}, \eta_{1:K}, \beta, \theta_{1:K} \right)$. See the Supplementary Material for mild  conditions  that ensure  convergence. 
\end{enumerate}

Given  
$(X_i^{(r)},W_i^{(r)},T_i^{(r)})_{i\ge1}$, $(X_i^{(e)},W_i^{(e)})_{i\ge1}$ and  the limit of        $\frac{n^{(e)}}{n^{(r+e)}}$, 
the vector $\theta^\circ_{1:K}$ can be represented  as a function of the parameters $\left( \nu_{1:K}, \eta_{1:K}, \beta, \delta_{1:K} \right)$ in model \eqref{model:logistic_true}. For instance, when $\delta_{1:K} = 0_{K \times 1}$ then the working model \eqref{model:logistic_mle} is correctly specified and   $\theta^\circ_{1:K} = \theta_{1:K}$  by standard maximum likelihood arguments. For other values of $\delta_{1:K}$ close to $0_{K \times 1}$ we have the Taylor approximation
\begin{equation} \label{logistic_taylor}
    \theta^\circ_{1:K} = \theta_{1:K} + B \delta_{1:K} + r_1 \left( \delta_{1:K}  \right),
\end{equation}
where $B =  \frac{\partial \theta_{1:K}^\circ}{\partial \delta_{1:K}}$ is the Jacobian matrix, and  
the remainder  $r_1 $ 
satisfies 
\begin{equation} \label{r_1_logistic}
    \lim_{ \delta_{1:K}  \to 0_{K \times 1}} \frac{ r_1 \left( \delta_{1:K}  \right)}{ || \delta_{1:K} ||_1} =  0_{K \times 1}.  
\end{equation}
We approximate the difference $\theta^\circ_{1:K} - \theta_{1:K}$ with a linear transformation of $\delta_{1:K}$. 
The approximation  is  analogous to equation \eqref{linear_r_e_bias} in subsection \ref{subsection:linear_model}.  
To select $\Sigma$ (in equation \ref{harmonized_simple}) and compute the harmonized estimator $\hat\theta_{1:K}^h$ we will use the factorization 
\begin{equation} \label{B_logistic}
B = \frac{\partial \theta_{1:K}^\circ}{\partial(\nu_{1:K}^\circ, \eta_{1:K}^\circ, \beta^\circ)}  \frac{\partial (\nu_{1:K}^\circ, \eta_{1:K}^\circ, \beta^\circ)}{\partial \delta_{1:K}}.
\end{equation}
Analytic expressions for computing the derivatives are included in the Supplementary Material.

The next result, similar to Proposition \ref{prop:rid_1}, focuses on  $\hat{\theta}^h_{1:K}$ when the sample sizes diverge. Here, $\hat{\theta}^h_{1:K}$
is computed using a data-dependent matrix $\hat{\Sigma}$, the value assigned to the regularization parameter  $\Sigma$ in equation \eqref{harmonized_simple}. 
As before,  $u = c \frac{\lambda}{\lambda + c} \Sigma \pi$, with  $c = \left( \pi^\top \Sigma \pi \right)^{-1}$.
\begin{proposition} \label{prop:harm_logistic_bias}  
      If assumption  \ref{assumption:logistic_n} 
      holds, and 
      the sequence of designs ensures  that  for any value of  $\delta_{1:k}$ in model \eqref{model:logistic_true}  
    \begin{enumerate}[wide = 0.5\parindent, leftmargin = 0.5\parindent, label = (B\arabic*)]

        \setcounter{enumi}{2}
        
        \item \label{assumption:r_asymptotic_unbiased} \hspace{0.1em} $\hat{\theta}^{(r)} \overset{p}{\to} \pi^\top \theta_{1:K}$,
        
        \item \label{assumption:re_asymptotic_linear_bias} \hspace{0.1em} $\hat{\theta}^{(r+e)}_{1:K} \overset{p}{\to} \theta_{1:K} + B \delta_{1:K} + r_1 \left( \delta_{1:K} \right)$, where   $r_1 \left( \delta_{1:K} \right) $ satisfies equation \eqref{r_1_logistic}, and
        
        \item \label{assumption:sigma_hat_converge} \hspace{0.1em} $\hat{\Sigma} \overset{p}{\to} \Sigma$, where $\Sigma$ is a positive-definite $K \times K$ matrix, 
    \end{enumerate} 
    then
    \begin{equation*}
        \hat{\theta}^h_{1:K} \overset{p}{\to} \theta_{1:K} + \left( I_K - u \pi^\top \right) B \delta_{1:K} +  r_2 \left(  \delta_{1:K} \right),
    \end{equation*}
    where the approximation error  $r_2 \left(  \delta_{1:K} \right)$ satisfies  
    \begin{equation*} \label{r_2_logistic}
        \lim_{ \delta_{1:K}  \to 0_{K \times 1}} \frac{ r_2 \left( \delta_{1:K}  \right)}{ || \delta_{1:K} ||_1} =  0_{K \times 1}.  
    \end{equation*}
\end{proposition}

In other words,  the bias of $\hat{\theta}^h_{1:K}$ is approximately equal to $\left( I_K - u \pi^\top \right) B \delta_{1:K}$. This allows us to select $\lambda$ and $\Sigma$ so that the harmonized estimator 
$\hat{\theta}^h_{1:K}$ is nearly unbiased 
under the SDM assumption \eqref{systematic_distortion_sec_3_2}.
In particular, when $ B 1_{K \times 1} \neq 0_{K \times 1}$, the estimator $\hat{\theta}^h_{1:K}$ with parameters $(\lambda,\Sigma)$ is approximately unbiased if (i) $\lambda = \infty$ and (ii)  $\Sigma \pi = \kappa  B 1_{K \times 1}$ for some $\kappa \neq 0$. In practice, we can compute a consistent estimate $\hat B$ of $B$, and select $\hat{\Sigma}$ such that 
\begin{equation} \label{logistic_Sigma}
    \hat{\Sigma} \pi = \kappa \hat{B} 1_{K \times 1}.
\end{equation}
It is only necessary that $\pi^\top \hat{B} 1_{K \times 1} \neq 0$ (see Proposition \ref{prop:sigma_existence}).
 With $\lambda =\infty$  all the positive matrices $\hat{\Sigma}$ that solve equation \eqref{logistic_Sigma} produce the same harmonized estimator. Note that if we fix $\kappa$, then
 the difference  between  the convex set  of solutions of equation $\eqref{logistic_Sigma}$
 and the  solutions of     ${\Sigma} \pi = \kappa {B} 1_{K \times 1}$ vanishes when the sample sizes increase.
To estimate $B$  we first compute the MLEs  $\left( \hat{\nu}_{1:K}^{(r)}, \hat{\eta}_{1:K}^{(r)}, \hat{\beta}^{(r)} \right)$ of model \eqref{model:logistic_mle} without using EC data and then plug these estimates into  the right-hand side of expression \eqref{B_logistic}. This does not require  Monte Carlo simulations or other approximation methods (see the Supplementary Material). The analytic computation of $\hat B$  utilizes the joint empirical distributions of covariates $X$ and biomarkers $W$ in the RCT and EC datasets. 

Although  here we  focused on logistic regression, the  approach to choose  $\hat \Sigma$ (see equations \ref{harmonized_simple} and \ref{logistic_Sigma}) that we described can also be used in other settings. 
 Consider, for example, count outcome data $Y$. We can proceed as follows:
 \vspace{-0.5em}
\begin{enumerate}[(a)]
    \item We specify a working model  --- say, a Poisson model with assumptions and covariate-outcome relationships similar to  the GLM  in expression  \eqref{model:logistic_mle} --- that leverages the EC and RCT data, to compute $\hat{\theta}^{(r+e)}_{1:K}$. We also compute 
      a consistent estimator $\hat{\theta}^{(r)}$ of the overall  treatment effect  $\theta$.
    \item  We specify  a  data-generating model, say an extended Poisson model
    with subgroup-specific bias terms  $\delta_{1:K}$ having the same interpretation as in expression \eqref{model:logistic_true}. 
    Under mild assumptions  (e.g., {\it iid} replicates of biomarkers, covariates and treatment in the RCT and external group)  the asymptotic difference $\theta_{1:K}^\circ - \theta_{1:K}$ is well defined.
    This difference can be viewed as a function of   $\delta_{1:K}$, and  is null if $\delta_{1:K}= 0_{K \times 1}$. 
%
    \item Therefore, as before, we can use the Taylor approximation $\theta_{1:K}^\circ - \theta_{1:K} \approx B \delta_{1:K}$, compute an estimate $\hat{B}$ of $B$, 
    find a positive definite matrix $\hat{\Sigma}$  that solves $\hat{\Sigma} =  \hat{B} 1_{K \times 1}$, and specify $\Sigma = \hat \Sigma$ in equation \eqref{harmonized_simple}. 
\end{enumerate} 
In this Poisson example, and in  other GLMs, $\hat{B}$ can be computed analytically using a nearly identical formula to the one that we use for logistic regression (see the Supplementary Material).

There are other strategies to select $\Sigma$. 
For example, as suggested by the Bayesian interpretation in Section \ref{subsection:cut}, one may choose $\Sigma$ to be an estimate of the variance of $\hat{\theta}_{1:K}^{(r+e)}$, giving a VD harmonized estimator. In our implementations we estimate this variance using  the Fisher information  of $\left( \nu_{1:K}, \eta_{1:K}, \beta \right)$ under  model \eqref{model:logistic_mle}.

Figure 5 illustrates a simulation study. We generate outcome data from model \eqref{model:logistic_true}, with $\delta_{1:K} = (\delta, \dots, \delta)$ for $\delta \in [-1, 1]$. This is a scenario with systematic bias across subgroups (SDM assumption holds); the Supplementary Material includes a scenario with bias terms $\delta_{1:K}$ that vary across subgroups.
We consider $K = 5$ subgroups.   
We generate a single covariate $X_i^{(r)} \overset{iid}{\sim} N(0, 1)$ in the RCT and $X_i^{(e)} \overset{iid}{\sim} N(2, 1)$ in the EC, with the regression coefficient $\beta = 0.2$.  The other  details  of the simulations are similar to the scenarios used for Figure \ref{fig:linear_density}. 
We compare four estimators of $\theta_{1:K}$: 
(i) the pooled estimator $\hat{\theta}_{1:K}^{(r+e)}$, using MLE based on model  \eqref{model:logistic_mle}; 
(ii) an RCT-only estimator denoted $\hat{\theta}^{(r)}_{1:K}$, using MLEs based on model \eqref{model:logistic_mle}
without the EC data; 
(iii) the BD harmonized estimator denoted $\hat{\theta}^{h-B}_{1:K}$, with $\lambda = \infty$ and $\Sigma$ selected using equation \eqref{logistic_Sigma}; and 
(iv) the VD harmonized estimator denoted $\hat{\theta}^{h-V}_{1:K}$, with $\lambda = \infty$ and $\Sigma$ equal to the  estimate of the variance of  $\hat{\theta}^{(r+e)}_{1:K}$. 
Panel A shows that the bias of $\hat{\theta}^{(r+e)}_{1}$ is  approximately linear in $\delta_1$, and  based on this fact harmonization can reduce the bias of $\hat{\theta}^{(r+e)}_{1}$ under the SDM assumption (see equation \eqref{systematic_distortion_sec_3_2}). Panel B shows that the relative performance of the estimators depends on $\delta_1$. Panel C shows that harmonization with $\Sigma$ chosen to reduce bias (BD harmonization) or  $\Sigma$ chosen as the estimated variance of $\hat{\theta}^{(r+e)}_{1:K}$ (VD harmonization) produce similar estimates in the simulation scenarios that we considered. Supplementary Figure S5  illustrates scenarios in which the SDM assumption is violated ($\delta_{1:K} = (\delta -0.5, \phantom{i} \delta + 0.5, \phantom{i} \delta - 0.5, \phantom{i} \delta + 0.5, \phantom{i} \delta - 0.5)$ with $\delta \in [-1,1]$) and the harmonized estimator presents lower RMSE than the RCT-only estimator despite post-harmonization bias.

\begin{figure}[tb!]
    \centering
    \includegraphics[width = 0.95\textwidth]{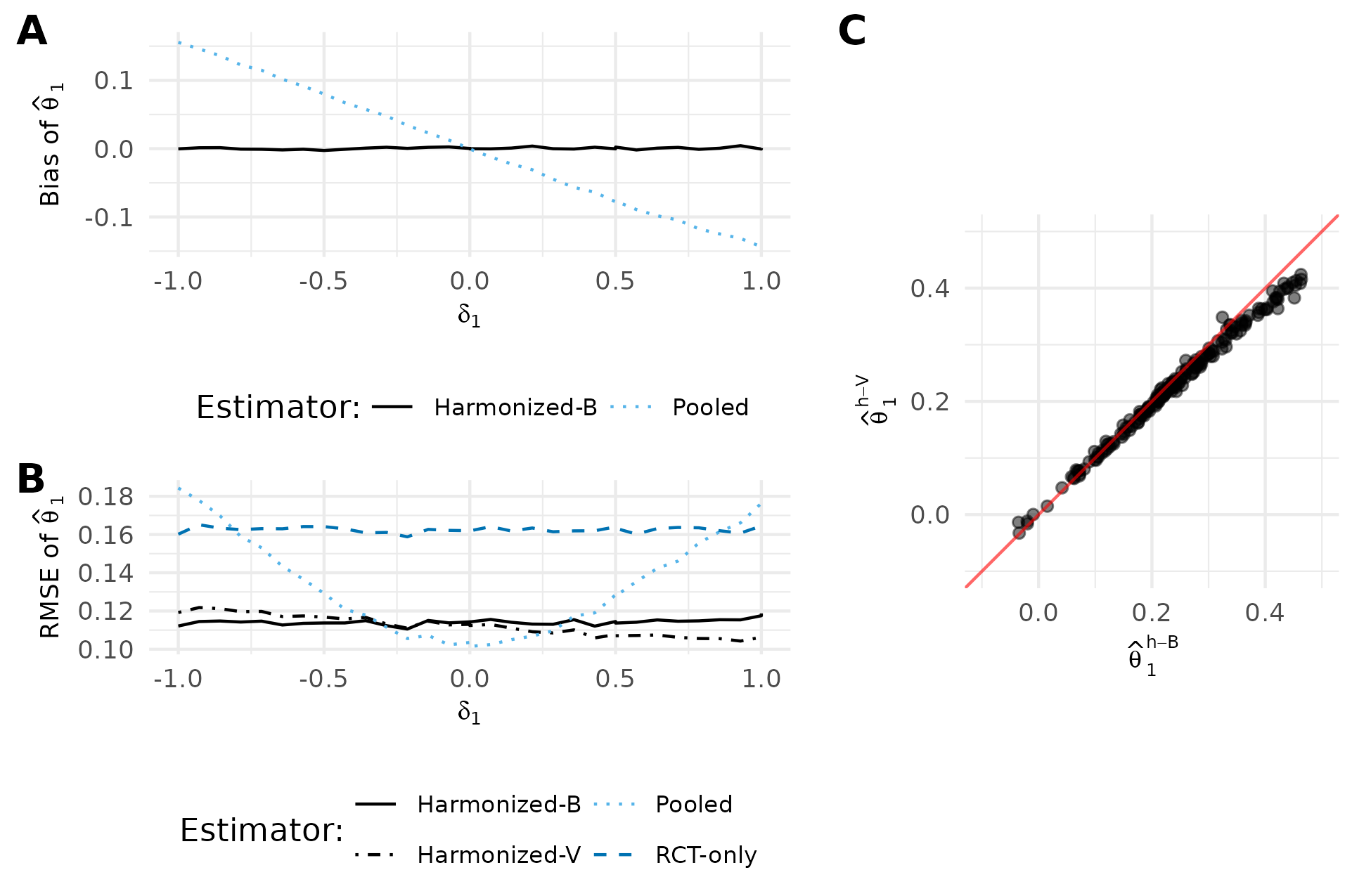}
    \caption{Logistic regression and harmonized estimates. 
    The results  are based on 2,000 replicates per scenario.
    (A) Bias when we estimate $\theta_1$  using the pooled estimator and the BD harmonized estimator as a function of $\delta_1$. 
    (B) The root mean squared error of  four estimators.
    (C) Scatter plot comparing the BD harmonized estimator (denoted $\hat{\theta}_1^{h-BD})$ and the VD harmonized estimator (denoted $\hat{\theta}_1^{h-VD})$ across a random subsample of 200 replicates with $\delta = 1_{K \times 1}$ and $\theta_1 = 0.23$.}
    \label{fig:logistic_fig}
\end{figure}

\nopagebreak[4]

\subsection{Propensity score weighting and harmonization} \label{subsection:ipw_logistic}

Statistical methods from the causal inference literature \citep{li_generalizing_2022}
can be useful for combining RCT and EC data.
We discuss the harmonization (equation \eqref{harmonized_def})  of subgroup-specific effect estimates obtained through propensity score weighting to leverage RCT and EC data. In Section \ref{section:GBM} we use this approach to analyze GBM data. 

Here we consider  
  an initial estimate $\hat{\theta}^{(r+e)}_{1:K}$ based on an inverse probability weighting scheme \citep{lunceford_stratification_2004, tipton_improving_2013}. 
Other approaches, such as marginal structural models \citep{cole_constructing_2008} and direct standardization  \citep{robins_new_1986, snowden_implementation_2011}, can  be used. 
We refer to \cite{li_generalizing_2022} and other reviews for  discussions of the principles, assumptions, and sensitivity of relevant causal inference methods.   
 We   consider again  binary outcomes, 
with the subgroup-specific treatment effects defined in equation \eqref{logistic_theta_k} and the overall treatment effect defined as $\theta = \pi^\top \theta_{1:K}$. 

To compute $\hat{\theta}^{(r+e)}_{1:K}$ we fit the logistic regression model \eqref{model:logistic_mle}, 
using weighted maximum likelihood, and apply equation \eqref{logistic_theta_k_re} as in subsection \ref{subsection:logistic_model} to estimate treatment effects. The individual log-likelihood contribution is multiplied by a weight $w_i^{(r)}$ or $w_i^{(e)}$, for RCT and EC patients respectively. In our implementation the weights are $w_i^{(r)} = 1$ for RCT patients and 
\begin{equation} \label{ipw_w_e}
    w_i^{(e)} = \zeta \frac{\hat{\rho} \left( W_i^{(e)}, X_i^{(e)} \right)}{1 - \hat{\rho} \left( W_i^{(e)}, X_i^{(e)} \right)}
\end{equation}
for EC patients, where $\hat{\rho} \left( W_i^{(e)}, X_i^{(e)} \right)$ is an estimate of the propensity score that we compute through  standard logistic regression  as in \cite{tipton_improving_2013}. 
The  constant $\zeta$ is selected  to have $\max_i w_i^{(e)} \le 1$, i.e.  the RCT patients are weighted more than the
 the patients in the EC
group. 
The ratio $\frac{\hat{\rho} \left( W, X\right)}{1 - \hat{\rho} \left( W, X \right)}$  in equation \eqref{ipw_w_e} has a straightforward interpretation.    
Indeed, if we multiply it by $\frac{n^{(e)}}{n^{(r)}}$ 
we obtain an estimate of the ratio of 
the densities at  potential $(W,X)$ values in the RCT and
 EC populations.  

To harmonize $\hat{\theta}^{(r+e)}_{1:K}$ we  compute the  treatment effect estimate
$\hat{\theta}^{(r)}$, which is based    on RCT data only. 
We use $\lambda = \infty$, and $\Sigma$ is selected with
     minor changes to the
 procedure  discussed in subsection \ref{subsection:logistic_model}: 
\vspace{-0.5em}
\begin{enumerate}[(a)]
    \item The working model  to compute $\hat{\theta}^{(r+e)}_{1:K}$  remains   \eqref{model:logistic_mle}. 
    Recall that we assign weights \eqref{ipw_w_e} to the EC patients to 
    compute the estimates  $\left( \hat{\nu}_{1:K}^{(r+e)}, \hat{\eta}_{1:K}^{(r+e)}, \hat{\beta}^{(r+e)} \right)$.

    \item We 
    build on the same assumptions as in subsection  \ref{subsection:logistic_model}, including the
     data-generating model  \eqref{model:logistic_true}.  We  focus on the   parameterization  $\left( \hat{\nu}_{1:K}^{(r)}, \hat{\eta}_{1:K}^{(r)}, \hat{\beta}^{(r)} \right)$, with   unknown bias components $\delta_{1:K}$. 
    The estimates   $\left( \hat{\nu}_{1:K}^{(r)}, \hat{\eta}_{1:K}^{(r)}, \hat{\beta}^{(r)} \right)$  are  based on \textit{RCT-only}  regression.
    Under this data-generating model  $\theta^\circ_{1:K}$, the limit  of  $\hat{\theta}^{(r+e)}_{1:K}$, is a function of $\delta_{1:K}$. 
    
    The limit $\theta^\circ_{1:K}$ can be  computed for any  $ ({\nu}_{1:K}^{(r)}, {\eta}_{1:K}^{(r)},{\beta}^{(r)},\delta_{1:K})$ configuration by maximizing the  a weighted log-likelihood function 
     with $4\times n^{(r)}+2\times n^{(e)}$ terms:
  2   terms ($Y_i^{(e)}=0$ or $1$) for each 
    patient in the  EC  group 
    (weights: the individual  $w_i^{(e)}$ weights  and the conditional  probability  that $Y^{(e)}_i=0$ or $1$ are multiplied)
    and 4
    ($T_i^{(r)}=0$ or $1$, and $Y_i^{(r)}=0$ or $1$)  for each 
    patient in the
     RCT group (weights: 
  $w_i^{(r)}=1$,   the conditional  probabilities  that $Y^{(r)}_i=0$ or $1$,
      and the probability of $T^{(r)}_i=0$ or $1$  are multiplied).  This approach assumes that the  distributions  of biomarkers and covariates $(X,W)$
        in the RCT and EC groups match the available empirical estimates.


    \item  Lastly  we use the Taylor approximation $\theta_{1:K}^\circ - \theta_{1:K} \approx B \delta_{1:K}$, where $B = \frac{\partial \theta^\circ_{1:K}}{\partial \delta_{1:K}}$. 
    We compute $\hat B$ through a weighted version of the analytic expressions that we used in subsection \ref{subsection:logistic_model}
    (see Supplementary Material). 
    As before, we select a matrix $\Sigma$ that solves the equation $\Sigma \pi = \kappa \hat B 1_{K \times 1}$. 
    Recall that  with $\lambda = \infty$ any  $\Sigma$  that solves the equation (i.e. BD harmonization) defines the same harmonized estimator.
    
\end{enumerate}
Alternatively, the analyst can  choose $\Sigma$  equal to  the estimate of the variance of $ \hat \theta ^{(r+e)}_{1:K}$, using popular sandwich estimators \citep{lunceford_stratification_2004} or  bootstrap procedures. 
See Supplementary Figure S3 (Section S2.8) for similar comparisons of 
    these estimators that focus on 
    the bias, variance, and RMSE when  
    the trial size $n^{(r)} = 100$ is fixed (with $n^{(r)}_{.,0} = 33$  and  $n^{(r)}_{.,1} = 67$), the EC sample size $n^{(e)}$ increases from $33$ to $627$ (i.e., $q = n^{(e)}_{.,0}(n^{(e)}_{.,0} + n^{(r)}_{.,0})^{-1}$ increases from 0.5 to 0.95) and the SDM assumption is violated. The bias and RMSEs of both the pooled and harmonized estimators increase with $n^{(e)} $. As expected this trend is less marked for the harmonized estimator compared to the pooled estimator. Importantly, when $n^{(e)}$ is large other definitions of $\hat{\theta}^{(r+e)}_{1:K}$, beyond pooling (e.g., Bayesian estimators based on hierarchical borrowing or power priors), can be used to reduce the outlined risks associated with distortion mechanisms that violate  the SDM assumption. 

\section{Analysis of glioblastoma data} \label{section:GBM}

We apply the harmonized estimator described in subsection \ref{subsection:ipw_logistic} to a glioblastoma (GBM) trial.
We include {\it real world} data in our analyses \citep{ventz_design_2019}.
The data collection that we use has been previously described \citep{rahman_accessible_2023}.
We evaluate the harmonized estimator through a model-free resampling scheme \citep{avalos-pacheco_validation_2023, ventz_use_2022}. The scheme is similar to bootstrapping and it  allows us to generate {\it in silico} RCTs and EC datasets. 
Using this resampling scheme we can explore the operating characteristics of different methods and subgroup analyses.

Subgroup analysis is an important task in GBM trials. Pre-clinical data often suggest that the effects of experimental  treatments might vary substantially across patient subgroups.  Also, experimental treatments might benefit only  some groups. See \cite{trippa_bayesian_2017} and references therein for an extensive discussion. As an early example, we mention  the  \cite{stupp_radiotherapy_2005} Phase III trial. The final analyses of  this large trial  provided evidence that temozolomide combined with  radiation therapy (TMZ+RT), which is the current standard of care in GBM, improves  survival  compared to radiation therapy (RT) alone. Subsequent subgroup analyses \citep{stupp_effects_2009} supported the hypothesis that the treatment benefit of TMZ+RT compared to RT  was   limited to patients with MGMT-negative methylation status.

{\it Data-collection:}
In our analyses we used data from the phase III AVAGLIO trial (profiles and outcomes of 352 newly diagnosed GBM patients treated with TMZ+RT) and larger real world data set from Dana Farber Cancer Institute (including 532 newly diagnosed GBM patients treated with TMZ+RT). We previously described these data collections in \cite{rahman_accessible_2023} and \cite{ventz_use_2022}.
We use individual-level patient data. We consider as the primary outcome 12-month  survival, i.e. $Y_i=1$ if the patient is alive 12 months after treatment initiation and $Y_i = 0$ otherwise. For patients  with a  follow-up  of less than 12 months  we imputed their 12-month survival status using a Cox model (see Supplementary Section S4.1 for details).
In our subgroup analyses we have $K=4$ non-overlapping subpopulations based on the patient's MGMT methylation status and their  Karnofsky performance status (KPS) before treatment. 
The four subgroups are: $(W_{i}=1)$ MGMT-negative and $KPS < 90$, $(W_{i}=2)$ MGMT-negative and $KPS \geq 90$, $(W_{i}=3)$ MGMT-positive and $KPS < 90$, and $(W_{i}=4)$ MGMT-positive and $KPS \geq 90$. The four subgroups have  proportions $(0.20, 0.47, 0.10, 0.23)$ in the trial and $(0.27, 0.30, 0.19, 0.24)$ 
in the real world dataset respectively. Relevant  pre-treatment covariates used in our analyses include age, sex, and extent of tumor resection prior to the  treatment. See \cite{ventz_design_2019} and references therein for a discussion on relevant prognostic variables in GBM.

{\it Resampling.}
We use a resampling procedure  \citep{ ventz_use_2022, avalos-pacheco_validation_2023} 
similar to bootstrapping
to generate {\it in silico} GBM RCTs.
Specifically, we generate $1,000$ {\it in silico} trials  repeating the following steps:
\vspace{-0.5em}
\begin{enumerate}[(a), wide = 0.5\parindent, leftmargin = 0.5\parindent]
    \item \textit{Control arm}. Sample $n^{(r)}_{\cdot,0} = 100$  
    patient records $\left( Y_{i}^{(r)}, W_{i}^{(r)}, X_{i}^{(r)} \right)$ with replacement from the actual  control arm of the clinical trial.

    \item \textit{Experimental arm}. Sample $n^{(r)}_{\cdot,1} = 200$ patient  records $\left( Y_{i}^{(r)}, W_{i}^{(r)}, X_{i}^{(r)} \right)$ with replacement again from the actual control arm of the clinical trial. 
    
    \item \textit{Treatment effects}.   Since we sampled   the same  group of  patients treated with TMZ+RT to generate the \textit{in silico} control  and experimental arms,  in our computer experiments the treatment effects are $\theta_{1:K} = (0, 0, 0, 0)$. Positive treatment effects can be spiked in by extending the survival  in the  \textit{in silico}  experimental arm  \citep{ventz_design_2019}.
    
    \item \textit{External data.} Sample $n^{(e)}_{\cdot, 0} = 600$ patient records  $\left( Y_{i}^{(e)}, W_{i}^{(e)}, X_{i}^{(e)} \right)$ from the actual EC dataset.

    \item \textit{Estimation}. Compute the subgroup-specific treatment effect estimates.  
    We compare  four    estimators:
    \vspace{-0.5em}
    \begin{enumerate}[(i)]
    
        \item The \textit{pooled} logistic regression estimator  
        (subsection \ref{subsection:logistic_model}, equation \ref{logistic_theta_k_re}).
        
        \item The propensity score \textit{weighted} logistic regression estimator from subsection \ref{subsection:ipw_logistic}, with weights defined in equation \eqref{ipw_w_e}.
        
        \item The \textit{harmonized} estimator. We harmonize 
        the weighted estimator  as proposed in subsection \ref{subsection:ipw_logistic}. We use $\lambda = \infty$ and specify $\Sigma$ equal to the identity matrix (see equation \ref{harmonized_simple}).
        
        \item An \textit{RCT-only} estimator  based on  maximum likelihood under model \eqref{model:logistic_mle}, without EC data.
    \end{enumerate}
\end{enumerate}
We chose $n^{(r)}_{.,0} = 100$, $n^{(r)}_{.,1} = 200$, and $n^{(e)} = 600$, realistic sample sizes close to the sizes of the source data sets.

\begin{figure}[tb!]
    \centering
    \includegraphics[width = 0.95\textwidth]{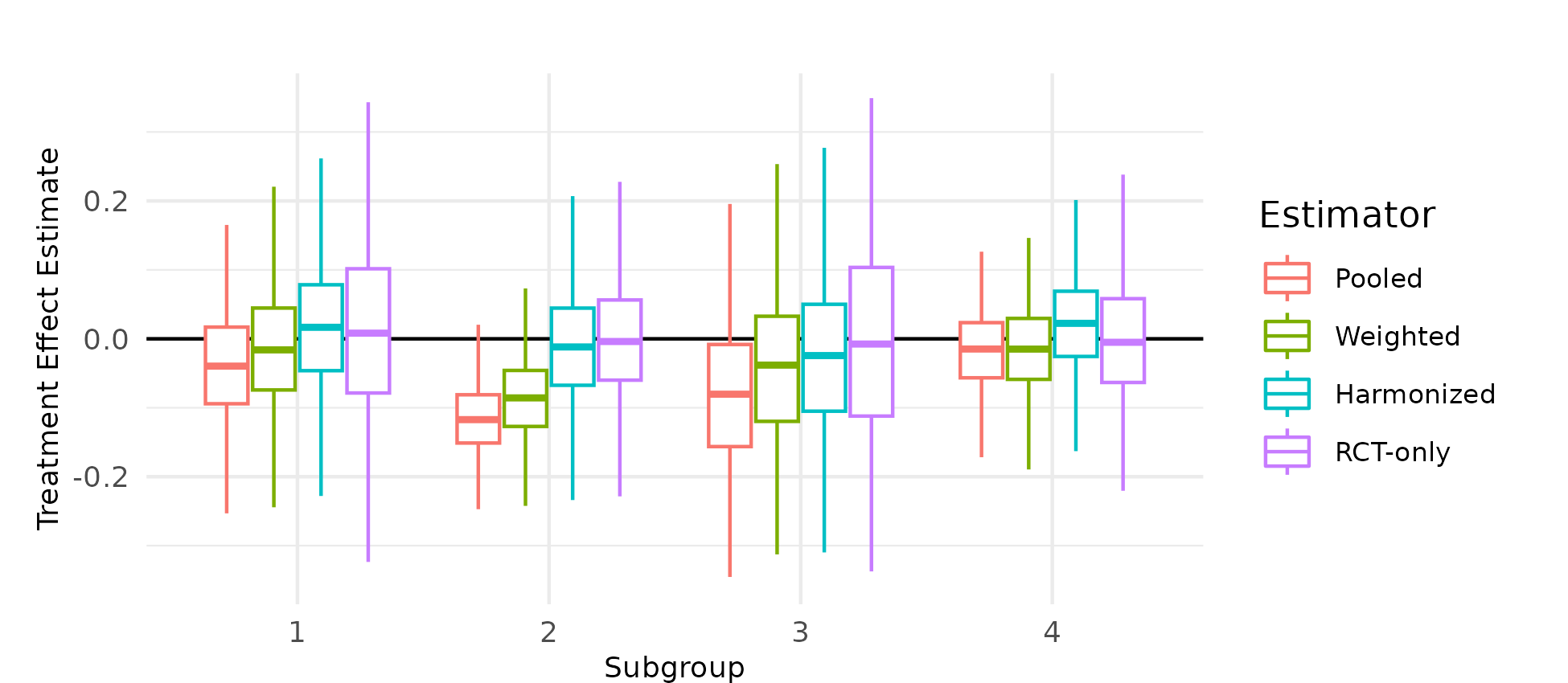}
    \caption{ The boxplots
     illustrate the distribution of the $\theta_{1:K}$ estimates across the 1,000 \textit{in silico} GBM trials simulated as we described.}
    \label{fig:gbm_fig}
\end{figure}

Figure \ref{fig:gbm_fig} shows the  distribution of the $\theta_{1:K}$ estimates across the \textit{in silico} RCTs. In all four subgroups the pooled estimator is the most biased. The propensity score weighted estimator slightly reduces this bias, and harmonization further reduces the  bias. 
We repeated the analyses 
implementing the data-dependent selection of  
$\Sigma$ to reduce bias as discussed in Section \ref{subsection:ipw_logistic}. 
The bias, compared to harmonization with fixed $\Sigma$ equal to the identity matrix, was reduced in 3 of the 4 subgroups, with reductions up to 90\%, while it remained nearly identical in one of the subgroups.
In addition, the propensity score weighted and harmonized estimators have similar variability. 
In particular, the standard deviations of the weighted and harmonized estimators are, respectively, 0.092 and 0.097 in subgroup 1, 0.060 and 0.081 in subgroup 2, 0.125 and 0.122 in subgroup 3, and 0.065 and 0.073 in subgroup 4. 
Harmonization in our computer experiments produces estimates with lower variability compared to the RCT-only estimator, which is unbiased. 
For further details on bias, variance, and RMSE in these simulations see Supplementary Table S1 (in Section S4.2).

\section{Discussion}

Subgroup analyses have become an essential component of testing the efficacy and safety of new experimental treatments in precision medicine. These analyses allow clinicians to identify subpopulations that respond to novel  therapies. However, conducting clinical trials  with sample sizes and rigorous statistical plans  for accurate  subgroup analyses is a challenging and costly task. Compared to RCTs designed to estimate the average treatment effect in the overall study population, trials designed to estimate subgroup-specific treatment effects require substantially larger sample sizes. 

These challenges may be partially addressed by allowing subgroup analyses to incorporate external control data from past trials and electronic health records. This can be particularly valuable for rare diseases and rare patient subpopulations. However, variations in the definitions of prognostic variables across studies, unmeasured confounders, different measurement standards across institutions, and other mechanisms can bias the research findings   when external datasets  augment RCT data.

In this work, we introduced  harmonized estimators for subgroup analysis that balance potential efficiency gains from the integration of EC data and the risk of bias due to confounding and other distortion mechanisms. The harmonized estimation procedure has two components: (i) an initial estimator $\hat \theta_{1:K}^{(r+e)}$ for the subgroup-specific treatment effects that leverages external control data and (ii) an estimator $\hat \theta^{(r)}$ for the average treatment effect in the RCT across subgroups that does not involve the use of external information. The proposed harmonization shrinks the initial subgroup-specific estimates $\hat \theta_{1:K}^{(r+e)}$ toward a linear subspace of potential subgroup-specific treatment effects with a prevalence-weighted average concordant  with the estimate  $\hat \theta^{(r)}$  obtained using only  the RCT  data. 
Any initial estimator $\hat{\theta}^{(r+e)}_{1:K}$ can be used and we provide strategies for selecting the parameters $\lambda$ and $\Sigma$. For example, throughout this paper we discuss parameterizations chosen with the goal of obtaining harmonized estimates $\hat{\theta}^h_{1:K}$ that are unbiased or approximately unbiased when the SDM assumption holds. In Section 2.1 we described how under some regularity conditions this bias-directed (BD) harmonized estimator can be viewed as a modified version of the  estimator $\hat{\theta}^{(r+e)}_{1:K}$ obtained by estimating and subtracting  the bias $\psi_{1:K} = E \left( \hat{\theta}^{(r+e)}_{1:K} \right) - \theta_{1:K}$ under the  SDM assumption. Similarly, BD harmonization of estimators $\hat \theta_{1:K}^{(r+e)}$ based on  GLMs (Section 3.2) subtracts from $\hat{\theta}^{(r+e)}_{1:K}$ an estimate of $\psi_{1:K}$. Compared to other strategies to reduce or remove bias due to SDMs  from subgroup analyses (see Supplementary Section S3.9 for an example), harmonization ensures coherence between subgroup analyses and  the primary analysis (i.e., $\pi^\top \hat{\theta}^h_{1:K} = \hat{\theta}^{(r)}$).

We discussed harmonized estimators 
based on various outcome models, including MLE for the linear and logistic regression models and procedures from the causal inference literature. We provided sufficient conditions to obtain BD harmonized treatment effect estimates. Using patient-level data  from brain cancer studies, we showed that, under realistic scenarios, the harmonized estimator can reduce the mean squared error of the subgroup-specific treatment effect estimates by up to 50\% compared to estimates that do not leverage external data. Moreover, our comparisons based on brain cancer trial data showed a reduction in bias via harmonization of up to 86\% compared to treatment effects estimates that use propensity score weighting to combine  external and RCT data. 

In our work we focused on harmonization with $\lambda = \infty$ to ensure that the harmonized estimates $\hat{\theta}^h_{1:K}$ satisfy $\pi^\top \hat{\theta}^h_{1:K} = \hat{\theta}^{(r)}$.  Moreover, taking $\lambda = \infty$ and selection of $\Sigma$ based on Proposition \ref{prop:rid_2} or equation \eqref{logistic_Sigma}, allows analysts to obtain a harmonized estimator that is unbiased  (or approximately unbiased) when the SDM assumption holds. Alternatively, taking $\Sigma$ as an estimate of the variance of $\hat{\theta}^{(r+e)}_{1:K}$ has a Bayesian justification.

Harmonization is a procedure applicable to any subgroup analysis that involves RCT data and ECs, including estimators $\hat{\theta}^{(r+e)}_{1:K}$ beyond those discussed  in this paper.  In Supplementary Section S3.7, we describe additional simulations in which harmonization improves MSE and reduces bias of subgroup-specific estimates $\hat{\theta}^{(r+e)}_{1:K}$ obtained  with (i) a Bayesian causal forest model \citep{hahn_bayesian_2020}, (ii) a hierarchical  model with borrowing of information across subgroups \citep{jones_bayesian_2011}, and (iii) a propensity score adjusted power prior procedure \citep{lin_propensityscorebased_2019}. In these simulations we considered the same scenarios as in Sections \ref{subsection:linear_model} and \ref{subsection:logistic_model}, and computed $\hat{\theta}^{(r+e)}_{1:K}$ (the posterior mean of $\theta_{1:K}$), which was then modified following the proposed harmonization approach. The comparisons of the operating characteristics pre- and post-harmonization remained similar to what we  illustrated throughout Section 3. In these cases it is challenging to analytically derive the bias of $\hat{\theta}^{(r+e)}_{1:K}$, but we can  compute BD harmonized estimators. This can be achieved by  estimating the bias of $\hat{\theta}^{(r+e)}_{1:K}$ with resampling  procedures (see Supplementary Section S3.10). We also point interested readers to Supplementary Section S5, where we discuss the bias of $\hat{\theta}^{(r+e)}_{1:K}$ and $\hat{\theta}^{h}_{1:K}$ in the presence of unmeasured predictors. These results show that various estimators can be harmonized, and that harmonization can improve their operating characteristics. In addition, harmonization can reduce the bias of $\hat\theta^{(r+e)}_{1:K}$ in scenarios in which the model used for estimating the treatment effects $\theta_{1:K}$ is misspecified.

\setstretch{1.2} 

\section*{Supplementary Material}
See the attached Supplementary Material file (PDF) for an extended notation dictionary, all derivations, and additional figures. 

\section*{Funding}
We gratefully acknowledge support from the National Institutes of Health under grants R01LM013352 (LT and DS), T32CA009337 (DS), and 5P30CA077598-25 (SV).

\begingroup
    \small
    \setlength{\bibsep}{5pt}
    \setstretch{1}
    \bibliography{RWD_Subgroup_Paper.bib} 
\endgroup

\end{document}